\documentclass[journal=apchd5,manuscript=article, layout=traditional]{achemso}

\usepackage{chemformula} % Formula subscripts using \ch{}
\usepackage[T1]{fontenc} % Use modern font encodings

\author{Maximilian Rödel}
\affiliation[University Würzburg, EP6]
{Lehrstuhl für Experimentelle Physik VI, Universität Würzburg, Am Hubland, 97074 Würzburg, Germany}
\email{maximilian.roedel@uni-wuerzburg.de}

\author{Polina Lisinetskaya}
\affiliation[University Würzburg, PhysTheoChem]
{Institut für Physikalische und Theoretische Chemie, Universität Würzburg, Am Hubland, 97074 Würzburg, Germany}

\author{Maximilian Rudloff}
\affiliation[University Würzburg, EP6]
{Lehrstuhl für Experimentelle Physik VI, Universität Würzburg, Am Hubland, 97074 Würzburg, Germany}

\author{Thomas Stark}
\affiliation[ZAE]
{Bayerisches Zentrum für Angewandte Energieforschung (ZAE Bayern e.V.), Magdalene-Schoch-Str. 3, 97074 Würzburg, Germany}

\author{Jochen Manara}
\affiliation[ZAE]
{Bayerisches Zentrum für Angewandte Energieforschung (ZAE Bayern e.V.), Magdalene-Schoch-Str. 3, 97074 Würzburg, Germany}

\author{Roland Mitric}
\affiliation[University Würzburg, PhysTheoChem]
{Institut für Physikalische und Theoretische Chemie, Universität Würzburg, Am Hubland, 97074 Würzburg, Germany}

\author{Jens Pflaum}
\affiliation[University Würzburg, EP6]
{Lehrstuhl für Experimentelle Physik VI, Universität Würzburg, Am Hubland, 97074 Würzburg, Germany}
\alsoaffiliation[ZAE]
{Bayerisches Zentrum für Angewandte Energieforschung (ZAE Bayern e.V.), Magdalene-Schoch-Str. 3, 97074 Würzburg, Germany}
\email{jpflaum@physik.uni-wuerzburg.de}

\title[Strong Coupling in Metal Organic Structures]
{The Role of Molecular Arrangement on the Dispersion in Strongly Coupled Metal-Organic Hybrid Structures}

\abbreviations{}

\begin{document}

%\begin{tocentry}

%\includegraphics[]{Deckblatt-Figure.pdf}

%Graphical table of contents, showing the main results of our study. The strong coupling with the crystalline phase and the intermediated layer which has a monolayer like character.

%\end{tocentry}

\begin{abstract}
Metal-organic hybrid structures have been demonstrated a versatile platform to study primary aspects of light-matter interaction by means of emerging states comprising excitonic and plasmonic properties. Here we are studying the wave-vector dependent photo-excitations in gold layers covered by molecular films of zinc-phthalocyanine and its fluorinated derivatives (F\textsubscript{n}ZnPc, with n = 0,4,8,16). These layered metal-organic samples show up to four anti-crossings in their dispersions correlating in energy with the respective degree of ZnPc fluorination. By means of complementary structural and theoretical data, we attribute the observed anti-crossings to three main scenarios of surface plasmon coupling: i) to aggregated $\alpha$ -phase regions within the F\textsubscript{n}ZnPc layers at 1.75 eV and 1.85 eV , ii) to a coexisting F\textsubscript{16}ZnPc $\beta$ -polymorph at 1.51 eV, and iii) to monomers, preferentially located at the metal interface, at 2.15 eV. Whereas energy and splitting of the monomer anti-crossings depend on strength and average tilting of the molecular dipole moments, the aggregate related anti-crossings show a distinct variation with degree of fluorination. These observations can be consistently explained by a change in F\textsubscript{n}ZnPc dipole density induced by an increased lattice spacing due to the larger molecular van der Waals radii upon fluorination. The reported results prove Au/F\textsubscript{n}ZnPc bilayers a model system to demonstrate the high sensitivity of exciton-plasmon coupling on the molecular alignment at microscopic length scales.
\end{abstract}

\section{Introduction}
The coupling between collective plasmonic excitations in metal thin films and localized excitations of a semiconductor in close proximity has become a topic of intense research in many areas of modern solid state physics \cite{PhysRevLett.93.036404,dintinger2005strong,liu2016strong}. The excitonic part of such hybrid materials can be realized, for instance, by inorganic semiconductors which benefit from the atomic precision in sample preparation by means of molecular beam epitaxy under ultra-high vacuum (UHV) conditions. As a result, semiconductor structures of different dimensionalities have been successfully demonstrated by quantum films, wells or dots. However, besides their rather narrow spectral variety and laborious processing under UHV and clean room conditions, III-V or II-VI semiconductors entail the problem of high dielectric constants yielding weak binding energies of their Wannier-type excitations and, thus, demand for cryogenic conditions. In contrast, low-weight carbon based molecules have small dielectric constants of about 3 to 4 and, hence, facilitate strong Coulomb bound electron-hole excitations of up to 1 eV binding energy\cite{hill2000charge}. This strong binding results in Frenkel-type excitations, typically localized on just one molecular site and being of remarkable stability with respect to interference with e.g. lattice phonons even at room temperature. Therefore, many interesting effects ranging from strongly enhanced absorption and emission \cite{kolb2017hybrid} to plasmon-polariton based lasers operating at room temperature \cite{zhu2017surface} have been reported for metal-organic hybrid structures in recent years.
Most of these innovative developments, however, rely on the coupling between local excitations of dyes and local plasmonic excitations of metallic nanoparticles and nanostructures or, alternatively, discrete cavity modes created by distributed Bragg reflectors \cite{lundt2016room, askitopoulos2011bragg}. Hence, the impressive versatility of such hybrid systems mainly results by changing the size and shape of the plasmonic material whereas the semiconducting component, especially its orientation in case of molecular thin films, is often considered to be of minor relevance and, as yet, has not been analyzed to the same extent. This becomes even more important as structuring of the metallic component, necessary for providing localized plasmonic excitations, and the resulting changes in the lateral surface polarizability will significantly affect the transition dipole moment alignment of the molecular entities deposited on-top. All together, these aspects have motivated us to present an alternative approach to such hybrid geometries which is based on delocalized surface plasmons in homogenous gold metal layers and their coupling to photo-excited states in the organic semiconductor F\textsubscript{n}ZnPc (n = 0,4,8,16) grown on top. The degree of fluorination enables access on the intermolecular interaction as well as the resulting molecular packing and, thereby, on the dispersion relation of the emerging quasi-particles being of coupled plasmonic and excitonic nature and often referred to as plexcitons. A prominent feature of plexcitonic states is their so-called anti-crossing (AC) which arises when the energy of the discrete semiconductor excitation intersects the surface plasmon polariton (SPP) dispersion. To lift the degeneracy of the two excitations, a splitting of the initial dispersion into two new branches, the upper and the lower plexciton band, occurs. 
In this contribution, by means of joint experimental and theoretical studies we investigate the  anti-crossings arising in the Au/F\textsubscript{n}ZnPc hybrid systems and trace them back to specific states originating by strong coupling. We demonstrate the plexciton dispersion as a sensitive probe for structural organization and intermolecular interaction in the organic semiconductor material.

\section{Experimental}

\begin{figure}
\includegraphics[]{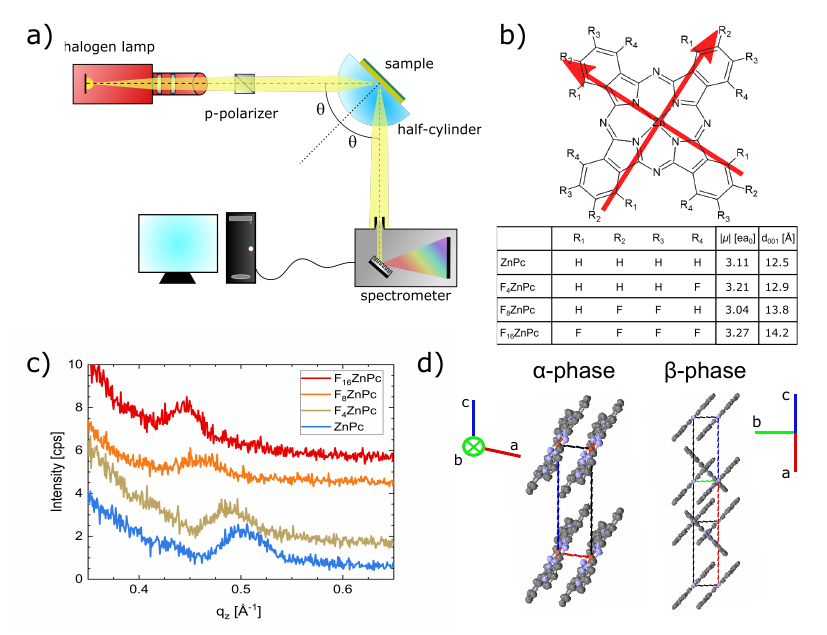}
\centering
\caption{a) Dispersion measurement setup operating in Kretschmann geometry. A p-polarized white light source is used for illumination of the sample in a $\theta $-2$\theta $ reflectivity configuration. b) Molecular structure of F\textsubscript{n}ZnPc with the respective side-group configuration, calculated transition dipole moments in atomic units as well as measured out-of-plane spacing d\textsubscript{001}. The red arrows exemplarily illustrate the transition dipole moments\cite{zhang2016visualizing}. c) X-ray diffraction measurements on ZnPc (blue), F\textsubscript{4}ZnPc (ochre), F\textsubscript{8}ZnPc (orange) and F\textsubscript{16}ZnPc (red) thin films. The observed peak refers to the (001) lattice spacing. d) $\alpha $- phase unit cell in case of ZnPc, F\textsubscript{4}ZnPc and F\textsubscript{8}ZnPc (left packing scheme) and unit cell of the $\beta $-phase polymorph for F\textsubscript{16}ZnPc (right packing scheme).}
\label{Abbildung1}
\end{figure}

The samples used in our study comprise a stacked layer geometry. Initially, pre-cleaned glass slides were covered by a 2 nm thick chromium wetting layer, followed by 50 nm gold film deposited via thermal evaporation in high vacuum (base pressure of $10^{-6} $ mbar) and constituting the plasmonic component of our hybrid structures. A self-assembled monolayer (SAM) of 1-Decanethiol chemisorbed on top of the gold prevents quenching of excitons near the metal interface. As excitonic counterpart, the organic semiconductor Zinc-Phthalocyanine (ZnPc) and its fluorinated derivatives F\textsubscript{4}-, F\textsubscript{8}- and F\textsubscript{16}ZnPc were chosen, the corresponding molecular structure is displayed in Figure \ref{Abbildung1}b). Besides the pronounced photostability, this compound enables a controlled energetic shift of the highest occupied and lowest unoccupied molecular orbital (HOMO and LUMO) upon fluorination without significantly affecting the related optical band gap. This unique property has rendered F\textsubscript{n}ZnPc an interesting candidate for opto-electronic applications like photovoltaics \cite{https://doi.org/10.1002/adfm.201404434, pfuetzner2011influence} or organic light emitting diodes \cite{hammer2019phase, van1996organic} and, as a result, has led to a comprehensive set of structural as well as optical data already available in literature \cite{OPITZ20091259, doi:10.1126/science.aaf0590}. Prior to sublimation, all organic materials were purified twice by gradient sublimation to avoid impurity effects on growth and optical properties. Two sample sets of sub-monolayer and 10 nm F\textsubscript{n}ZnPc layer thickness were prepared by molecular beam deposition under high vacuum (base pressure of $10^{-8} $ mbar). The chosen thicknesses provide insights in the photophysics related to the monomer as well as to the aggregate while simultaneously guaranteeing sufficient sensitivity on the interfacial light-matter coupling.\\
In a first set of studies, we confirmed the out-of-plane lattice spacing and, thereby, the molecular orientation of each F\textsubscript{n}ZnPc layer in the hybrid structures by X-ray diffraction. The corresponding (001) peaks measured in Bragg-Brentano geometry are shown in Figure \ref{Abbildung1}c) and indicate a decrease in the perpendicular momentum transfer $q_{z}$, i.e. an out-of-plane lattice expansion, with increasing degree of fluorination. Quantitative analysis of the peak positions yields the corresponding (001) lattice spacings d\textsubscript{001} for ZnPc, F\textsubscript{4}ZnPc, F\textsubscript{8}ZnPc and F\textsubscript{16}ZnPc as listed in Figure \ref{Abbildung1}b). According to previous studies, we attribute the out-of-plane lattice expansion together with the enhanced vertical alignment of the molecules to their larger van der Waals radii and hence inter-molecular repulsion upon fluorination \cite{https://doi.org/10.1002/adfm.201404434}. Moreover, our estimated (001) lattice spacings indicate the presence of the F\textsubscript{n}ZnPc $\alpha$-phase polymorph\cite{P.ErkCrystal} which has been observed only in thin films and is illustrated by the left unit cell in Figure \ref{Abbildung1}d) for the case of ZnPc\cite{berger2000studies}. As can be seen, the $\alpha$-phase is characterized by a pairwise arrangement of molecules within the unit cell which renders the lowest photoexcited state to be of excimeric nature. The wavefunction associated to this excited dimer is preferentially distributed over two next-neighbouring molecules and, thus, shows only a small oscillator strength depending on their respective distance. An exception to the stabilized thin film $\alpha$-phase is given by F\textsubscript{16}ZnPc\cite{doi:10.1021/ja064641r} which is known to form the bulk $\beta$-phase polymorph already at room temperature and is displayed on the right in Figure \ref{Abbildung1}d)\cite{https://doi.org/10.1002/adfm.201404434}. The latter becomes energetically more favorable, presumably, due to the stronger repulsive intermolecular interaction which at the same time destabilizes the $\alpha$-phase\cite{brown1968crystal}.

\section{Computational}

Theoretical simulations of optical properties of isolated F\textsubscript{n}ZnPc (n = 0,4,8,16) molecules were performed using linear- response TDDFT method with long-range and dispersion-corrected functional $\omega $B97X-D \cite{chai2008long} and 6-311++G** basis set for all atoms as implemented in the Gaussian16 package\cite{g16}. The molecular geometries were optimized, and electronic excitation energies and corresponding transition dipole moments were obtained for the two lowest excited states. The transition dipole moments are schematically presented in Figure \ref{Abbildung1}b) and their magnitudes are summarized in the table below the schema.\\
In order to investigate the excitonic states which arise in ordered F\textsubscript{n}ZnPc molecular aggregates we employed our previously developed theoretical approach presented in  \cite{lisinetskaya2016first, roehr2016excitonic, lisinetskaya2019collective} in details and briefly described in the SI. The intermolecular coupling was described using the transition charge method \cite{madjet2006intermolecular} implemented as described in \cite{lisinetskaya2016first}. It allowed us to mimic the interaction between electron densities and thus, to take into  consideration the multipolar interaction higher than the dipole-dipole one. For densely packed molecular aggregates it is crucial to account for spatial charge distribution and the transition charge approach provides this possibility at reasonably low computational costs.
The plexcitonic dispersion was modelled by combining the Jaynes-Cummings model \cite{1443594} adopted to the situation of a spatially delocalized SPP coupled to a localized exciton \cite{PhysRevLett.110.126801, yuen2016plexciton, yuen2018molecular} with the previously determined excitonic Hamiltonian (see the SI for details).

\section{Optical Characterization}

\begin{figure}
\includegraphics[width=1.0\textwidth]{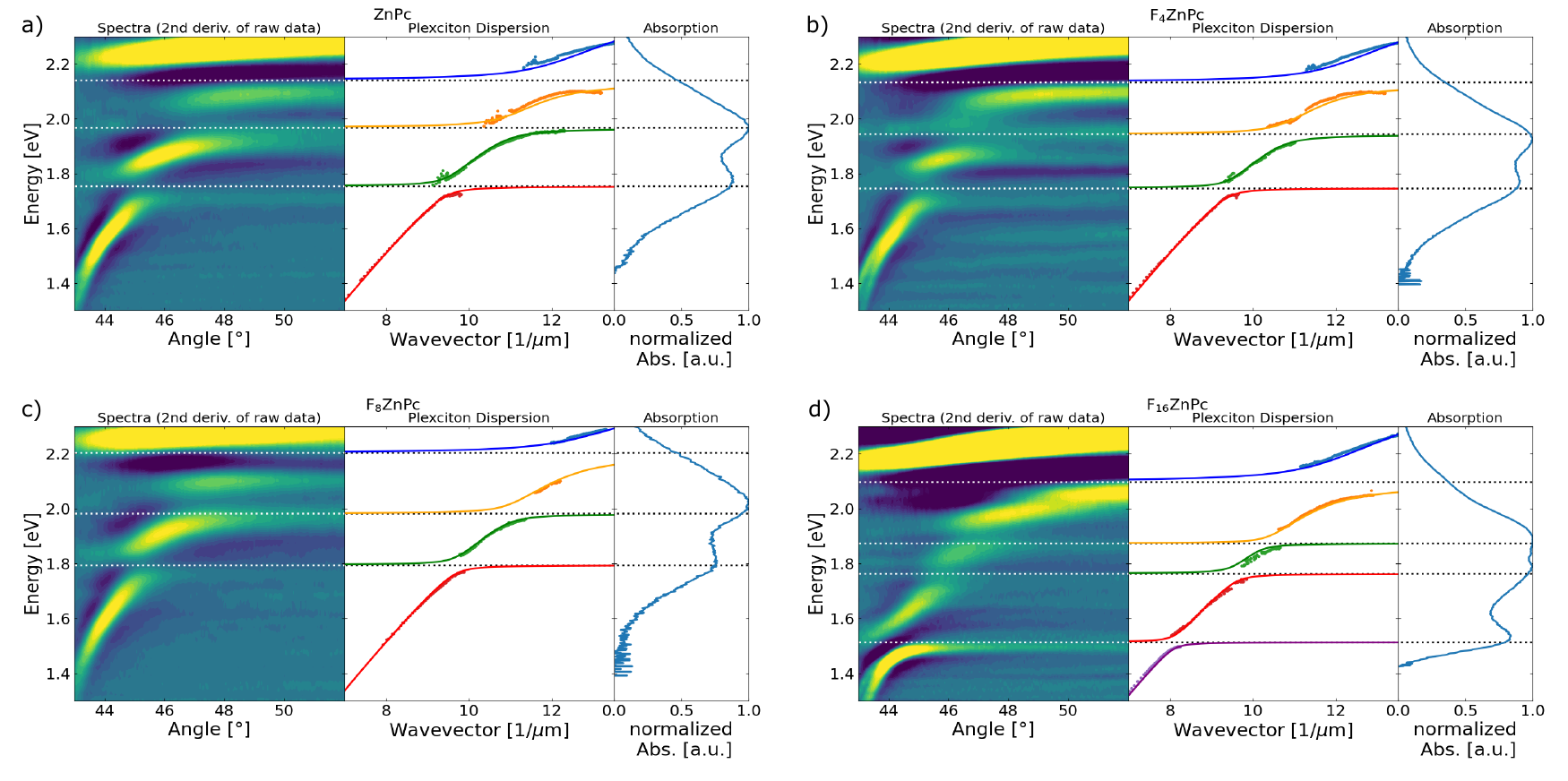}
\centering
\caption{Measured plexciton spectra for a) ZnPc, b) F\textsubscript{4}ZnPc, c) F\textsubscript{8}ZnPc and d) F\textsubscript{16}ZnPc, respectively. On the left side of each subfigure, the second derivative of the normalized raw data is displayed, which shows 3 to 4 anti-crossings for each material combination (dotted lines). In the middle graphs, the plexciton dispersions calculated from the peak positions are shown. On the right side, the normalized absorption spectra for each organic semiconductor is plotted. The exciton energies estimated from the plexciton dispersions (dotted lines in the corresponding plots), clearly match the peak positions in the absorption spectra and, thus, are unique for each material combination.}
\label{Abbildung2}
\end{figure}

\begin{figure}
\includegraphics[]{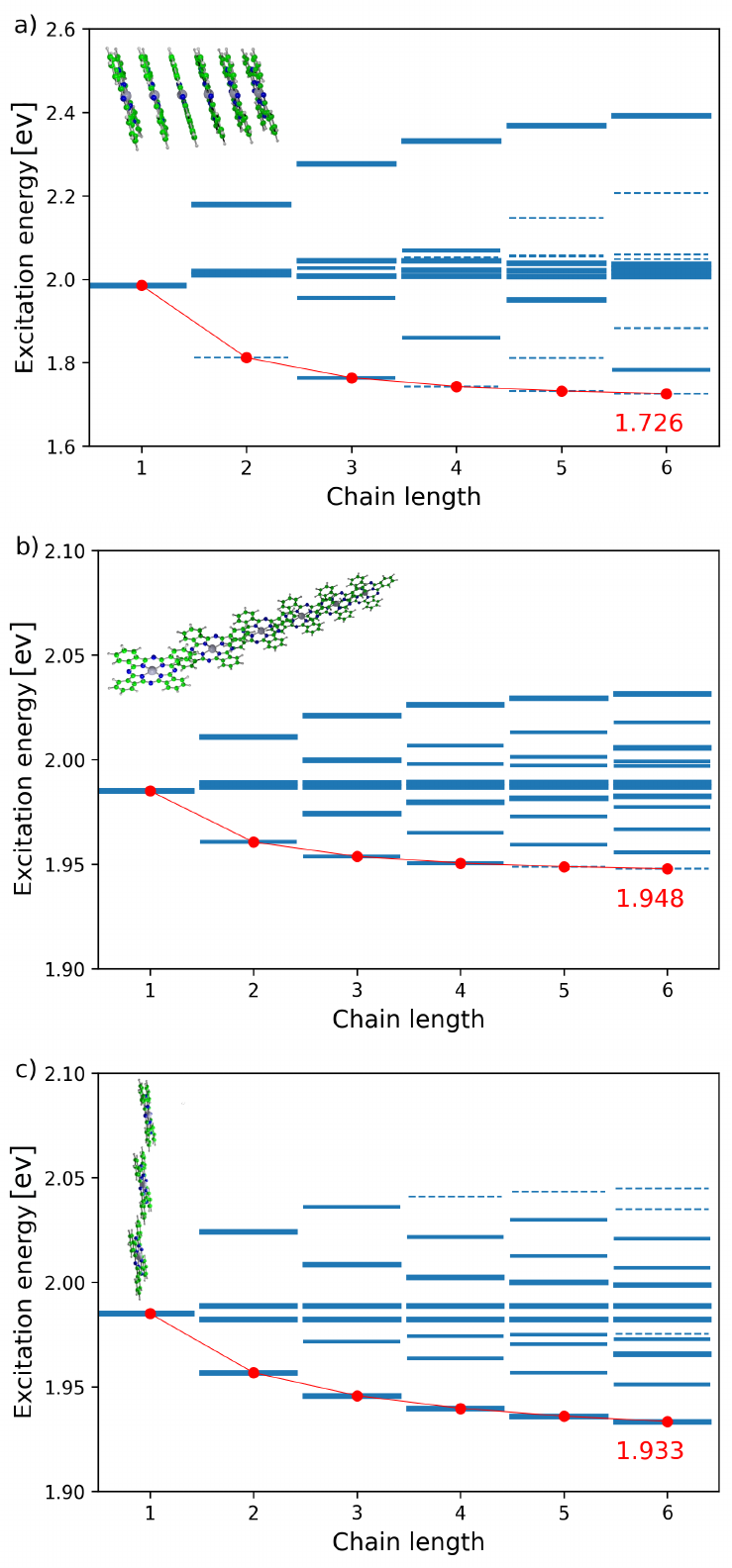}
\centering
\caption{Excitonic energies of 1-dimensional chains of ZnPc molecules packed in the direction of a) unit cell vector $\vec{a}$, b) unit cell vector $\vec{b}$, and c) unit cell vector $\vec{c}$. The red dots mark the lowest excitonic state energy as a function of the chain length, the excitonic energy of a hexamer is provided in red. The chain configuration is shown in insets.}
\label{Abbildung3}
\end{figure}

\begin{figure}
\includegraphics[]{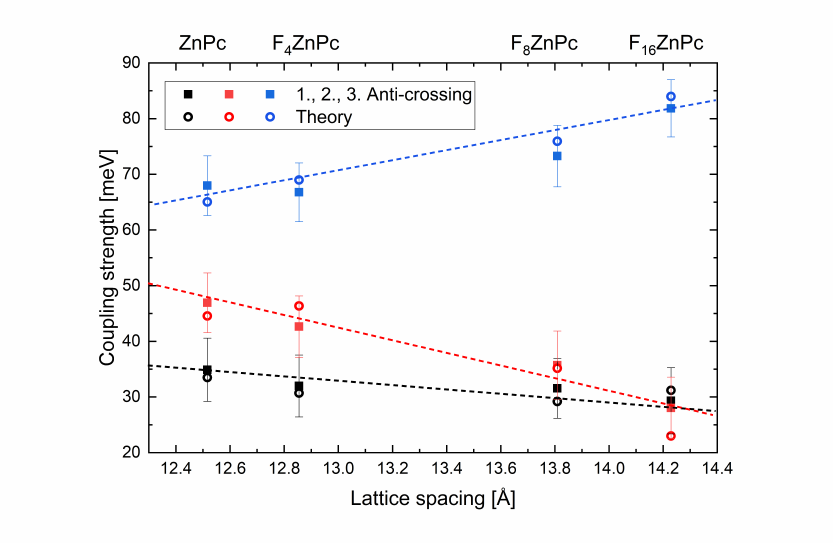}
\centering
\caption{Lattice spacing $d\textsubscript{001}$ and experimental(squares) as well as theoretical(circles) coupling strength $V$ for ZnPc, F\textsubscript{4}ZnPc, F\textsubscript{8}ZnPc and F\textsubscript{16}ZnPc, respectively. Due to the repulsive forces between the fluorinated ligands of adjacent molecules, the intermolecular interaction changes, leading to a larger out-of- plane lattice spacing with increasing degree of fluorination. Overall, this results in a less dense packing with increasing fluorination, consequently weakening the coupling between excitonic states and the surface plasmon polariton compared to the $\alpha $-phase regions (anti-crossings 2 and 3).}
\label{Abbildung4}
\end{figure}

\begin{figure}
\includegraphics[width=1.0\textwidth]{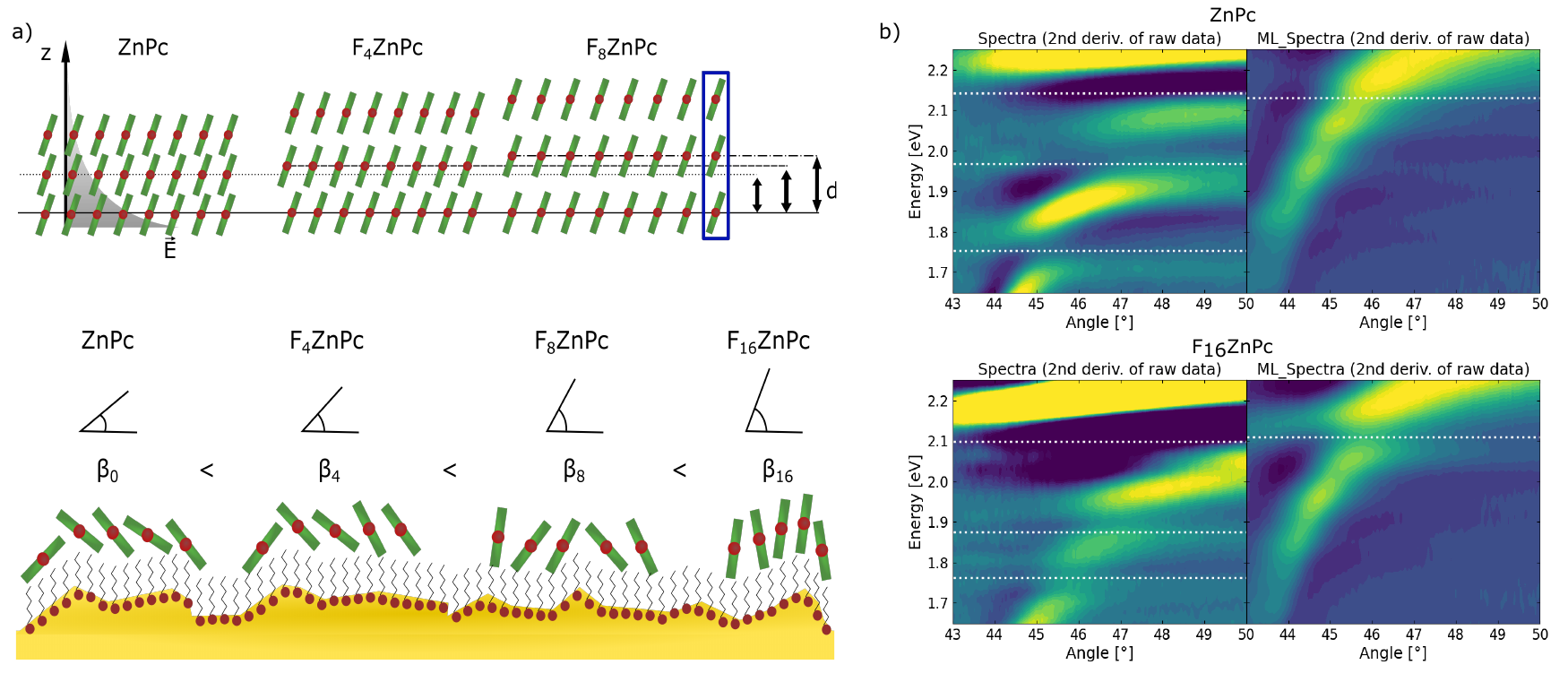}
\centering
\caption{a) Stacking of differently fluorinated F\textsubscript{n}ZnPc crystallites. Upon fluorination the out-of-plane lattice spacing of the $\alpha $-phase increases, leading to smaller coupling constant due to the decrease of emitter density. The blue frame highlights the 1D emitter arrangement. Below, the molecular packing in the interfacial region is illustrated. By its monomer-type photophysical behavior and increase in inclination angle $\beta\textsubscript{n}$ this disordered region leads to excitons of high energy causing the highest lying anti-crossing in our plexciton dispersions in Figure \ref{Abbildung2}. b) Plexciton dispersions of sub-monolayer thick ZnPc and F\textsubscript{16}ZnPc grown on top of SAM-functionalized Au metal films (right side) together with the dispersions of the corresponding 10nm thick films for comparison (left side). The coincidence of the high energetic anti-crossing at 2.1 eV is indicates its monomeric origin.}
\label{Abbildung5}
\end{figure}

Aiming for the plexcitons, their dispersions and, in particular, anti-crossings ermerging upon coupling between metal substrate and organic coverlayer, we employed an optical goniometer to illuminate the hybrid thin films in Kretschmann geometry as illustrated in Figure \ref{Abbildung1}a). P-polarized white light of a collimated halogen lamp is guided on the sample. To guarantee an achromatic alignment of the beam after incidence and reflection, a glass half cylinder is used as dispersive medium. The sample is placed on the planar backside of the half cylinder with an index matching oil and the reflected light is detected by a spectrometer. Normalizing the reflected light intensity in this $\theta $-2$\theta $ arrangement (s. fig. \ref{Abbildung1}a)), we are able to identify  resonantly excited plexcitons by their specific absorption characteristics in the spectra.\\
Figure \ref{Abbildung2} compiles the observed dispersion curves of the four different Au (50 nm)/SAM/F\textsubscript{n}ZnPc (10nm) samples. To improve the signal contrast, the second derivative of the raw intensity data is displayed on the left side of each subfigure a)-d). The bright yellow regions correspond to the absorption dip caused by the photoexcitation of the exciton-plasmon polariton. As can be seen for all samples, the dispersion at small energies and momenta, i.e. small angles, is characterized by a linear increase, which is indicative for the plasmonic part of the excitation. Towards larger angles, the dispersion curves approximate a horizontal trend and becomes independent of the momentum, which refers to the nature of the exciton and its energy. According to these intensity maps, each Au/SAM/F\textsubscript{n}ZnPc hybrid sample shows three anti-crossings at approximately 1.75 eV, 1.95 eV and 2.10 eV, indicated by the dashed horizontal lines in Figure \ref{Abbildung2}, except for F\textsubscript{16}ZnPc which shows an additional anti-crossing located at lower energy of 1.51 eV. Modelling the experimental data by an appropriate number of Lorentzians, i.e. one for each individual dispersion branch, and extracting the related peak position as function of energy and momentum results in the plexciton dispersions shown in the middle graphs of subfigures \ref{Abbildung2}a)-d). Fitting, in a final step, the obtained energy-momentum-relations by equation (S17) (see the SI) we can determine the energy of each participating excitonic state, $E$, and the related coupling strength $V$. Both values are listed in Table \ref{Table1} for each Au/SAM/F\textsubscript{n}ZnPc hybrid structure and labelled in an ascending order with respect to the energy.\\
\begin{table}
  \caption{Exciton energy and coupling strengths extracted from fitting the dispersion relations in Figure \ref{Abbildung2} a)-d), respectively. A qualitative trend of lower splitting strength with increasing F\textsubscript{n}ZnPc fluorination can be deduced for the low energy anti-crossings. For the highest energy anti-crossing this trend is reversed.}
  \label{Table1}
\begin{tabular}{|c|c|c|c|c|c|c|c|c|}
\hline 
 & \multicolumn{2}{c|}{0. anti-crossing} & \multicolumn{2}{c|}{1. anti-crossing} & \multicolumn{2}{c|}{2. anti-crossing}& \multicolumn{2}{c|}{3. anti-crossing}\tabularnewline
\hline 
 & E (eV) & V (meV) & E (eV) & V (meV) & E (eV) & V (meV)& E (eV) & V (meV)\tabularnewline
\hline 
\hline 
ZnPc &  &  & 1.754 & 35 & 1.968 & 47 & 2.141 & 68\tabularnewline
\hline 
F$_{4}$ZnPc &  &  & 1.749 & 32 & 1.944 & 43 & 2.137 & 67\tabularnewline
\hline 
F$_{8}$ZnPc &  &  & 1.795 & 32 & 1.983 & 36 & 2.180 & 73\tabularnewline
\hline 
F$_{16}$ZnPc & 1.513 & 25 & 1.763 & 29 & 1.874 & 28 & 2.098 & 82\tabularnewline
\hline 
\end{tabular}
\end{table}
In order to correlate these excitonic states with the optical characteristics of the respective organic semiconductor, we measured the absorption spectra of neat 10 nm thick F\textsubscript{n}ZnPc films on glass, displayed on the right side of each subfigure \ref{Abbildung2}a)-d). Obviously, for all bilayer hybrid samples the two anti-crossings located at around 1.75 eV and 1.95 eV coincide with the optical absorption transitions in the organic semiconductor component. More precisely, these transitions can be attributed to the spectral Q-band signature of the ZnPc $\alpha $-phase, which is characterized by these two absorption maxima\cite{ELNAHASS2004491,WOJDYLA20063441}. Accordingly, the associated anti-crossings refer to the coupling of the metal surface plasmon polariton to the excitations of the F\textsubscript{n}ZnPc $\alpha $-phase aggregate and thus, represent the excitonic contribution of the crystalline regions within the organic layer.\\
This assignment is supported by the theoretical simulations of excitonic states in the ordered chains of F\textsubscript{n}ZnPc molecules. Three types of molecular packing were investigated, namely, packing in the direction of the shortest unit cell vector ($\vec{a}$ in Figure \ref{Abbildung1}d)), unit cell vectors $\vec{b}$ and $\vec{c}$. The simulations reveal, that the excitonic states at around 1.75 eV are indeed characteristic to ordered densely packed F\textsubscript{n}ZnPc molecular structures ($\vec{a}$-direction), and the excitonic states at appr. 1.95 eV are observed along more sparse molecular chains ($\vec{b}$-direction and $\vec{c}$-direction). In the 3-dimensional molecular structures both sets of excitonic states are present giving rise to the 1st and 2nd anti-crossings observed in the experiment. The excitonic energies for ZnPc chains up to hexamer are presented in Figure \ref{Abbildung3}, for the fluorinated analogues see the SI.\\
Furthermore, comparing the spectral absorption of the F\textsubscript{16}ZnPc thin films with the dispersion of the corresponding metal-organic hybrid layer it becomes obvious that the anti-crossing at lowest energy correlates with the optical transition at 1.51 eV. As both features exclusively appear for F\textsubscript{16}ZnPc and as it is known from previous studies that this derivative forms a coexisting $\beta $-phase/$\beta_{bilayer} $-phase polymorph already at room temperature we relate these spectral characteristics to the same morphological origin \cite{https://doi.org/10.1002/adfm.201404434, OPITZ20091259}. 
Remarkably, whereas in the X-ray diffraction pattern shown in Figure \ref{Abbildung1}c) no signature of the coexisting $\beta $-phase \cite{hosokai2010simultaneous} can be resolved, the optical absorption as well as plexciton dispersion clearly reveal its presence and thereby, impressively demonstrate their superior sensitivity on molecular length scales. Nevertheless, reality might be somewhat more complex as the $\beta $-phase can be initiated by a bilayer precursor phase of several monolayer thickness depending on the substrate \cite{doi:10.1021/ja064641r}. This F\textsubscript{16}ZnPc $\beta_{bilayer} $-phase shows an $\alpha $-phase like stacking and, by its proximity to the gold metal surface and significantly higher volume fraction at this thickness range, might provide the main contribution to the dispersion taking into account the exponential decay of the SPP electric field strength. Vice versa, the neat $\beta $-volume phase is located further apart from the metal-organic interface and hence, contributes only by a small fraction to the overall signal. Moreover, as structural investigations on the F\textsubscript{16}CuPc analogue reveal that the transformation between $\beta_{bilayer} $ and $\beta $-volume phase can proceed over more than ten monolayers, we might face the situation that our F\textsubscript{16}ZnPc thin film sample of about seven monolayers ($\approx $10 nm) thickness is not completely converted but rather represents a mixture of $\beta $-phase layers with different packings and spacings \cite{doi:10.1021/ja064641r, hosokai2010simultaneous}.\\
The larger van der Waals radius and stronger repulsion between the per-fluorinated peripheries are considered the main reason for the $\alpha $- to $\beta $-structural phase transformation in F\textsubscript{16}ZnPc, which directly raises the question how the gradual fluorination of our compounds, either by induced structural or electronic changes, will affect the splitting and hence, coupling in the anti-crossing regions. For this purpose, the coupling strength, $V$, together with the energetic position of the related anti-crossing, $E_{exc}$, is listed in Table \ref{Table1}. Figure \ref{Abbildung4} shows the relation between coupling strength of the three highest anti-crossings occurring in all hybrid samples and (001) lattice spacing of the corresponding F\textsubscript{n}ZnPc layer determined by X-ray analysis (s. Figure \ref{Abbildung1}b).\\
This comparative illustration elucidates remarkable aspects and trends in the coupling behavior. At first, the overall high coupling strengths of approximately 40 meV up to 75 meV for anti-crossings 1, 2 and 3, respectively, are indicative for the large dipole moment of the molecular constituents as calculated above. We clearly can distinguish two different trends in the variation of coupling strength with the degree of fluorination. Whereas the coupling at anti-crossing points 1 and 2 declines or remains almost constant in strength with increasing fluorination, anti-crossing 3 shows apparently the reversed behavior, i.e. an increase in coupling strength upon fluorination. Accordingly, these opposite tendencies in the coupling have to be assigned to structural reasons, as the variation of the dipole strength between the different compounds is too small (s. fig. \ref{Abbildung1}b)) to account for the observed effects. Based on these calculations and the similar magnitudes of the F\textsubscript{n}ZnPc transition dipole moments we can further conclude that the almost twice as large coupling strength at anti-crossing point 3 hints at a closer distance of this molecular species to the metal interface compared to those being the origin of anti-crossings 1 and 2. As will be discussed below, this result matches perfectly the assumption of a rather disordered molecular monolayer located directly at the metal surface and, thus, resembling monomeric character in its optical properties.\\
Let us first focus on the behavior of anti-crossings 1 and 2. Since our metal-organic hybrid samples are of identical stratification and nominal layer thicknesses, the main structural difference is the lattice spacing of the F\textsubscript{n}ZnPc top-layer and, associated with that, the emitter density we had introduced by equations (S13) and (S14) in the SI. However, as the angle between evanescent plasmonic field $\vec{E}$ and molecular dipole $\vec{\mu}$ remains almost unchanged and therewith, the inherent coupling constant $V\sim\vec{E}\cdot\vec{\mu} $, we have to correlate the lattice spacing with the intermolecular packing and the resulting emitter density. For this purpose, we quantify the packing within our F\textsubscript{n}ZnPc layers by their 1D molecular density $\rho $ along the surface normal of the samples as indicated in Figure \ref{Abbildung5}a).
Calculating the relative 1D F\textsubscript{n}ZnPc emitter density with respect to that of ZnPc we are able to consistently describe the relative variations in coupling strength at the corresponding anti-crossings. As can be seen by Table \ref{Table2}, the 1D density ratios are a direct measure for the ratios of the associated coupling strength of each material. Hence, we conclude the emitter density to be the main origin for the variation in exciton-plasmon polariton interaction of our sample series. In a next step, we extend this model to a 3D emitter density by assuming similar in-plane and out-of-plane lattice constants as for the F\textsubscript{n}CuPc analogues reported in literature \cite{P.ErkCrystal,pandey2012resolving,jiang2017molecular}. In this case, the $\vec{a}$-direction of the reduced unit-cell remains nearly constant while $\vec{b}$- and $\vec{c}$-direction expand. Assuming a fixed spacing along the $\vec{a}$-axis and a similar relative expansion for the $\vec{b}$- and $\vec{c}$-direction yields a decrease in emitter density compared to the 1D case (see Table \ref{Table2} for comparison).\\
\begin{table}
 \caption{Calculated 1D/3D emitter density and coupling strength ratio for each F\textsubscript{n}ZnPc analogs.}
 \label{Table2}
\begin{tabular}{|c|c|c|c|c|c|c|c|c|}
\hline 
 & \multicolumn{2}{c|}{ZnPc} & \multicolumn{2}{c|}{F\textsubscript{4}ZnPc} & \multicolumn{2}{c|}{F\textsubscript{8}ZnPc}& \multicolumn{2}{c|}{F\textsubscript{16}ZnPc}\tabularnewline
\hline 
 & 1. AC & 2. AC & 1. AC & 2. AC & 1. AC & 2. AC & 1. AC & 2. AC\tabularnewline
\hline 
\hline 
$V_{ZnPc}/V_{F_{n}ZnPc} $ & 1 & 1 & 1.09 & 1.10 & 1.11 & 1.31 & 1.19 & 1.67\tabularnewline
\hline 
1D: $\sqrt{\rho_{ZnPc}}/\sqrt{\rho_{F_{n}ZnPc}} $ &  \multicolumn{2}{c|}{1}  & \multicolumn{2}{c|}{1.01} & \multicolumn{2}{c|}{1.05} & \multicolumn{2}{c|}{1.07}\tabularnewline
\hline 
3D: $\sqrt{\rho_{ZnPc}}/\sqrt{\rho_{F_{n}ZnPc}} $ &  \multicolumn{2}{c|}{1}  & \multicolumn{2}{c|}{1.03} & \multicolumn{2}{c|}{1.10} & \multicolumn{2}{c|}{1.14}\tabularnewline
\hline 
\end{tabular}
\end{table}
The theoretical calculations of the energies and coupling strengths at the anti-crossings were carried out for molecular hexamers packed in $\vec{a}$-, $\vec{b}$-, and $\vec{c}$-direction. The results are summarized in Table S1 in the SI. The energetic positions of the anti-crossings as well as their variation with the degree of fluorination are in a very good agreement with the experimental data. The trends experimentally observed in the coupling strengths are reproduced overall as well. Nevertheless, one can recognize that the calculated coupling values are generally lower for the 1\textsuperscript{st} anti-crossing and higher for the 2\textsuperscript{nd} one compared to the experimental ones. This can be attributed to the excitonic model used to simulate the coupled molecular aggregate, which underestimates transition dipole moments for the chains packed in the $\vec{a}$-direction (H-aggregate-like) and overestimates these for the chains packed in the $\vec{b}$- and $\vec{c}$-directions (J-aggregate-like, cf. insets in Figure \ref{Abbildung3}). Another reason for overestimation of the coupling strengths could be the closer proximity to the metal surface in our theoretical considerations. In our real samples, due to the interaction with the SAM-functionalized gold support, the molecules will not directly start stacking in the $\alpha $-phase but will form some intermediate transition layer compensating for substrate induced effects by an increased disorder and distance of molecular entities to the surface, overall, lowering the effective coupling strength. Actually, this proves to be a valid assumption, as we shall see in later discussions. In Figure \ref{Abbildung4} both datasets from theory and experiment are depicted, after correcting the former for a coupling offset for better comparison. Both anti-crossings are showing same trends, where anti-crossing 1 has a rather small decline (ca. 15\%) and anti-crossing 2 a greater decrease (ca. 40\%) upon gradual fluorination.
We have to mention that in order to optain a full data set for each degree of fluorination we used crystallographic data for F\textsubscript{n}CuPc which is considered equivalent in scientific discussions \cite{P.ErkCrystal, pandey2012resolving, jiang2017molecular}.\\
Finally, we have to clarify the origin of the remaining high energy anti-crossing at about 2.15 eV occurring for all Au/SAM/F\textsubscript{n}ZnPc hybrid layers. Taking into account the almost identical molecular dipole moment for all compounds under study, the strong splitting at this anti-crossing has to be related to a strong amplitude of the evanescent electric field of the surface plasmon polariton and thus, to a region of the organic layer localized in close proximity to the metal surface \cite{PhysRevLett.110.126801}. We therefore assume this high energy anti-crossing to originate from the interfacial region of the F\textsubscript{n}ZnPc layers on top of the Au/SAM substrates. This assumption is corroborated by the finding that the energy of the corresponding absorption transition agrees well with that of the F\textsubscript{n}ZnPc monomer observed in either solution \cite{savolainen2008characterizing}, diluted films or calculated theoretically\cite{doi:10.1021/jp407608w} and thus, hints at a rather disordered molecular arrangement within this film fraction. Again, as the calculated energies for spatially anisotropic packed monomers in supplement Table S2  show the right trend, we suggest that the high polar environment, caused by the metal thin film, leads to a constant blueshift of the exciton energy by approximately  175 meV. Obviously, as indicated in the schematic drawing in Figure \ref{Abbildung5}a) the roughness of the gold substrate layer, being further enhanced by the self-assembled monolayer on top, leads to a less defined orientation of molecules at the initial stage of the layer growth \cite{peisert2001order}. As reported for similar metal/Pc heterostructures, the molecular orientation can cover a wide range from horizontal to almost upright standing\cite{IKAME2005373,doi:10.1021/jp061889u,lozzi2003xps,peisert2001order} thus avoiding the formation of long range ordered aggregates of defined orientation as well as the related anti-crossings at lower energy.\\
To experimentally confirm this hypothesis, we prepared two additional sets of samples composed of ZnPc as well as F\textsubscript{16}ZnPc as organic top-layer, in this case however, with a sub-monolayer nominal thickness of 1 nm. The plexciton dispersions measured on these hybrid structures are plotted in Figure \ref{Abbildung5}b), showing the high energy anti-crossing to occur at 2.15 eV for Au/ZnPc as well as Au/F\textsubscript{16}ZnPc samples whereas the anti-crossings at lower energies, indicative for thicker crystalline films, are absent. Again, as for the shift in emitter density upon fluorination, this result underlines the high sensitivity of our method to even fractions of a molecular monolayer.\\
Comparing the coupling constants at the third anti-crossing for the four different compounds we observe an increase with gradual fluorination (Figure \ref{Abbildung4}). Theoretical calculations assuming a random oriention, however, show at first only minor effects by the degree of fluorination as can be seen by comparing the theoretical coupling constants of the third anti-crossing with the experiment (see Table \ref{Table1} and Table S2 I and II in the SI). For higher fluorination the coupling strength indeed is not changing significantly while in the experiments an increase is observed. Theory predicts, that the coupling constant is stronger along the z-direction, i.e. along the surface normal, than in the (xy)-plane, so that $V$ should be mainly determined by the averaged molecular inclination angle within the planes parallel to the substrate surface. For SAM-functionalized substrates the interaction with the adsorbed molecules is expected to be weak or only moderate compared to the bare metal surface, rendering the assumption of a preferred upright molecular orientation at the interface reasonable\cite{peisert2001order}. As a measure for this behavior, even in case of rather distorted molecular arrangement, a mean inclination angle $\beta_n$ (n = 0,4,8,16) with respect to the surface is introduced. Consistent with the explanation of an increase in lattice spacing with increasing fluorination, the change in intermolecular interaction will lead to larger mean inclinations as shown schematically in Figure \ref{Abbildung5}a) yielding an enhanced coupling strength as corroborated by the measured dispersions. Theoretical simulations performed for F\textsubscript{n}ZnPc sub-monolayers with gradually increasing inclination angle fully support this conclusion and again show the same trends like the experiments (see Figure \ref{Abbildung4}). Utilizing this high sensitivity of our dispersion curves on the individual molecular packing and alignment we can, to a large extent, reconstruct the morphology and, hence, the various stages of thin film growth in our Au/F\textsubscript{n}ZnPc hybrid layers with (sub-)monolayer resolution.

\section{Conclusion}

We demonstrated the strong light-matter interaction in F\textsubscript{n}ZnPc (n=0,4,8,16) thin films deposited on functionalized gold substrates as function of fluorination and thickness. By measuring the dispersion curves we were able to assign the energetic position of the anti-crossings to the absorption transitions in the respective organic layer. We attributed the shape and characteristics of these dispersions to two main contributions: The coupling of the aggregated F\textsubscript{n}ZnPc phase to the electric field of the surface-plasmon polariton (SPP), with the coupling constant being correlating to the particular lattice spacing of the molecular layer. The second contribution emerges by the SPP coupling to the interfacial layer of the molecular films which by its disordered molecular alignment proves to be of monomeric character in its optical properties. Due to its close proximity to the gold surface and by roughly estimating the number of molecules forming this intermediated region, we could substantiate that the size of the coupling strength for the different F\textsubscript{n}ZnPc layers on gold originates not only from the difference in emitter density but also from different mean angles of molecular orientation with respect to the z-component of the electric feld attending the surface plasmon polaritons. This finding proves plexciton dispersion a valuable tool for investigating the structural properties in metal-organic hybrid systems and renders the molecular arrangement an additional degree of freedom to tune the energetics of the photonic states emerging upon strong light-matter coupling.

\begin{acknowledgement}

M.R. and J.P. acknowledge financial support by the Bavarian State Ministry for Science and the Arts within the collaborative research network “Solar Technologies go Hybrid” (SolTech). We thank Peter Erk from BASF SE (RCS - J542S, 67056 Ludwigshafen am Rhein, Germany) for supplying part of the organic materials.

\end{acknowledgement}

\begin{suppinfo}

\begin{itemize}
\item SuppInfo.pdf: Supporting information for the article. Includes theoretical descriptions.
\end{itemize}

\end{suppinfo}

\bibliography{2021_UniWue_FnZnPc_coupling_ACSPhotonics}

\providecommand{\latin}[1]{#1}
\makeatletter
\providecommand{\doi}
  {\begingroup\let\do\@makeother\dospecials
  \catcode`\{=1 \catcode`\}=2 \doi@aux}
\providecommand{\doi@aux}[1]{\endgroup\texttt{#1}}
\makeatother
\providecommand*\mcitethebibliography{\thebibliography}
\csname @ifundefined\endcsname{endmcitethebibliography}
  {\let\endmcitethebibliography\endthebibliography}{}
\begin{mcitethebibliography}{41}
\providecommand*\natexlab[1]{#1}
\providecommand*\mciteSetBstSublistMode[1]{}
\providecommand*\mciteSetBstMaxWidthForm[2]{}
\providecommand*\mciteBstWouldAddEndPuncttrue
  {\def\EndOfBibitem{\unskip.}}
\providecommand*\mciteBstWouldAddEndPunctfalse
  {\let\EndOfBibitem\relax}
\providecommand*\mciteSetBstMidEndSepPunct[3]{}
\providecommand*\mciteSetBstSublistLabelBeginEnd[3]{}
\providecommand*\EndOfBibitem{}
\mciteSetBstSublistMode{f}
\mciteSetBstMaxWidthForm{subitem}{(\alph{mcitesubitemcount})}
\mciteSetBstSublistLabelBeginEnd
  {\mcitemaxwidthsubitemform\space}
  {\relax}
  {\relax}

\bibitem[Bellessa \latin{et~al.}(2004)Bellessa, Bonnand, Plenet, and
  Mugnier]{PhysRevLett.93.036404}
Bellessa,~J.; Bonnand,~C.; Plenet,~J.~C.; Mugnier,~J. Strong Coupling between
  Surface Plasmons and Excitons in an Organic Semiconductor. \emph{Phys. Rev.
  Lett.} \textbf{2004}, \emph{93}, 036404\relax
\mciteBstWouldAddEndPuncttrue
\mciteSetBstMidEndSepPunct{\mcitedefaultmidpunct}
{\mcitedefaultendpunct}{\mcitedefaultseppunct}\relax
\EndOfBibitem
\bibitem[Dintinger \latin{et~al.}(2005)Dintinger, Klein, Bustos, Barnes, and
  Ebbesen]{dintinger2005strong}
Dintinger,~J.; Klein,~S.; Bustos,~F.; Barnes,~W.~L.; Ebbesen,~T. Strong
  coupling between surface plasmon-polaritons and organic molecules in
  subwavelength hole arrays. \emph{Phys. Rev. B} \textbf{2005}, \emph{71},
  035424\relax
\mciteBstWouldAddEndPuncttrue
\mciteSetBstMidEndSepPunct{\mcitedefaultmidpunct}
{\mcitedefaultendpunct}{\mcitedefaultseppunct}\relax
\EndOfBibitem
\bibitem[Liu \latin{et~al.}(2016)Liu, Lee, Naylor, Ee, Park, Johnson, and
  Agarwal]{liu2016strong}
Liu,~W.; Lee,~B.; Naylor,~C.~H.; Ee,~H.-S.; Park,~J.; Johnson,~A.~C.;
  Agarwal,~R. Strong exciton--plasmon coupling in MoS2 coupled with plasmonic
  lattice. \emph{Nano Lett.} \textbf{2016}, \emph{16}, 1262--1269\relax
\mciteBstWouldAddEndPuncttrue
\mciteSetBstMidEndSepPunct{\mcitedefaultmidpunct}
{\mcitedefaultendpunct}{\mcitedefaultseppunct}\relax
\EndOfBibitem
\bibitem[Hill \latin{et~al.}(2000)Hill, Kahn, Soos, and
  Pascal~Jr]{hill2000charge}
Hill,~I.; Kahn,~A.; Soos,~Z.; Pascal~Jr,~R. Charge-separation energy in films
  of $\pi$-conjugated organic molecules. \emph{Chem. Phys. Lett.}
  \textbf{2000}, \emph{327}, 181--188\relax
\mciteBstWouldAddEndPuncttrue
\mciteSetBstMidEndSepPunct{\mcitedefaultmidpunct}
{\mcitedefaultendpunct}{\mcitedefaultseppunct}\relax
\EndOfBibitem
\bibitem[Kolb and Pflaum(2017)Kolb, and Pflaum]{kolb2017hybrid}
Kolb,~V.; Pflaum,~J. Hybrid metal-organic nanocavity arrays for efficient light
  out-coupling. \emph{Opt. Express} \textbf{2017}, \emph{25}, 6678--6689\relax
\mciteBstWouldAddEndPuncttrue
\mciteSetBstMidEndSepPunct{\mcitedefaultmidpunct}
{\mcitedefaultendpunct}{\mcitedefaultseppunct}\relax
\EndOfBibitem
\bibitem[Zhu \latin{et~al.}(2017)Zhu, Xu, Wang, Zhang, Deotare, Agrawal, and
  Lezec]{zhu2017surface}
Zhu,~W.; Xu,~T.; Wang,~H.; Zhang,~C.; Deotare,~P.~B.; Agrawal,~A.; Lezec,~H.~J.
  Surface plasmon polariton laser based on a metallic trench Fabry-Perot
  resonator. \emph{Sci. Adv.} \textbf{2017}, \emph{3}, e1700909\relax
\mciteBstWouldAddEndPuncttrue
\mciteSetBstMidEndSepPunct{\mcitedefaultmidpunct}
{\mcitedefaultendpunct}{\mcitedefaultseppunct}\relax
\EndOfBibitem
\bibitem[Lundt \latin{et~al.}(2016)Lundt, Klembt, Cherotchenko, Betzold, Iff,
  Nalitov, Klaas, Dietrich, Kavokin, H{\"o}fling, \latin{et~al.}
  others]{lundt2016room}
Lundt,~N.; Klembt,~S.; Cherotchenko,~E.; Betzold,~S.; Iff,~O.; Nalitov,~A.~V.;
  Klaas,~M.; Dietrich,~C.~P.; Kavokin,~A.~V.; H{\"o}fling,~S., \latin{et~al.}
  Room-temperature Tamm-plasmon exciton-polaritons with a WSe 2 monolayer.
  \emph{Nat. Commun.} \textbf{2016}, \emph{7}, 1--6\relax
\mciteBstWouldAddEndPuncttrue
\mciteSetBstMidEndSepPunct{\mcitedefaultmidpunct}
{\mcitedefaultendpunct}{\mcitedefaultseppunct}\relax
\EndOfBibitem
\bibitem[Askitopoulos \latin{et~al.}(2011)Askitopoulos, Mouchliadis, Iorsh,
  Christmann, Baumberg, Kaliteevski, Hatzopoulos, and
  Savvidis]{askitopoulos2011bragg}
Askitopoulos,~A.; Mouchliadis,~L.; Iorsh,~I.; Christmann,~G.; Baumberg,~J.;
  Kaliteevski,~M.; Hatzopoulos,~Z.; Savvidis,~P. Bragg polaritons: strong
  coupling and amplification in an unfolded microcavity. \emph{Phys. Rev.
  Lett.} \textbf{2011}, \emph{106}, 076401\relax
\mciteBstWouldAddEndPuncttrue
\mciteSetBstMidEndSepPunct{\mcitedefaultmidpunct}
{\mcitedefaultendpunct}{\mcitedefaultseppunct}\relax
\EndOfBibitem
\bibitem[Brendel \latin{et~al.}(2015)Brendel, Krause, Steindamm, Topczak,
  Sundarraj, Erk, Höhla, Fruehauf, Koch, and
  Pflaum]{https://doi.org/10.1002/adfm.201404434}
Brendel,~M.; Krause,~S.; Steindamm,~A.; Topczak,~A.~K.; Sundarraj,~S.; Erk,~P.;
  Höhla,~S.; Fruehauf,~N.; Koch,~N.; Pflaum,~J. The Effect of Gradual
  Fluorination on the Properties of FnZnPc Thin Films and FnZnPc/C60 Bilayer
  Photovoltaic Cells. \emph{Adv. Funct. Mater.} \textbf{2015}, \emph{25},
  1565--1573\relax
\mciteBstWouldAddEndPuncttrue
\mciteSetBstMidEndSepPunct{\mcitedefaultmidpunct}
{\mcitedefaultendpunct}{\mcitedefaultseppunct}\relax
\EndOfBibitem
\bibitem[Pfuetzner \latin{et~al.}(2011)Pfuetzner, Mickel, Jankowski, Hein,
  Meiss, Schuenemann, Elschner, Levin, Rellinghaus, Leo, \latin{et~al.}
  others]{pfuetzner2011influence}
Pfuetzner,~S.; Mickel,~C.; Jankowski,~J.; Hein,~M.; Meiss,~J.; Schuenemann,~C.;
  Elschner,~C.; Levin,~A.~A.; Rellinghaus,~B.; Leo,~K., \latin{et~al.}  The
  influence of substrate heating on morphology and layer growth in C60: ZnPc
  bulk heterojunction solar cells. \emph{Org. Electron.} \textbf{2011},
  \emph{12}, 435--441\relax
\mciteBstWouldAddEndPuncttrue
\mciteSetBstMidEndSepPunct{\mcitedefaultmidpunct}
{\mcitedefaultendpunct}{\mcitedefaultseppunct}\relax
\EndOfBibitem
\bibitem[Hammer \latin{et~al.}(2019)Hammer, Ferschke, Eyb, and
  Pflaum]{hammer2019phase}
Hammer,~S.; Ferschke,~T.; Eyb,~G.~v.; Pflaum,~J. Phase transition induced
  spectral tuning of dual luminescent crystalline zinc-phthalocyanine thin
  films and OLEDs. \emph{Appl. Phys. Lett.} \textbf{2019}, \emph{115},
  263303\relax
\mciteBstWouldAddEndPuncttrue
\mciteSetBstMidEndSepPunct{\mcitedefaultmidpunct}
{\mcitedefaultendpunct}{\mcitedefaultseppunct}\relax
\EndOfBibitem
\bibitem[Van~Slyke \latin{et~al.}(1996)Van~Slyke, Chen, and
  Tang]{van1996organic}
Van~Slyke,~S.~A.; Chen,~C.; Tang,~C.~W. Organic electroluminescent devices with
  improved stability. \emph{Appl. Phys. Lett.} \textbf{1996}, \emph{69},
  2160--2162\relax
\mciteBstWouldAddEndPuncttrue
\mciteSetBstMidEndSepPunct{\mcitedefaultmidpunct}
{\mcitedefaultendpunct}{\mcitedefaultseppunct}\relax
\EndOfBibitem
\bibitem[Opitz \latin{et~al.}(2009)Opitz, Ecker, Wagner, Hinderhofer,
  Schreiber, Manara, Pflaum, and Brütting]{OPITZ20091259}
Opitz,~A.; Ecker,~B.; Wagner,~J.; Hinderhofer,~A.; Schreiber,~F.; Manara,~J.;
  Pflaum,~J.; Brütting,~W. Mixed crystalline films of co-evaporated hydrogen-
  and fluorine-terminated phthalocyanines and their application in photovoltaic
  devices. \emph{Org. Electron.} \textbf{2009}, \emph{10}, 1259--1267\relax
\mciteBstWouldAddEndPuncttrue
\mciteSetBstMidEndSepPunct{\mcitedefaultmidpunct}
{\mcitedefaultendpunct}{\mcitedefaultseppunct}\relax
\EndOfBibitem
\bibitem[Schwarze \latin{et~al.}(2016)Schwarze, Tress, Beyer, Gao, Scholz,
  Poelking, Ortstein, Günther, Kasemann, Andrienko, and
  Leo]{doi:10.1126/science.aaf0590}
Schwarze,~M.; Tress,~W.; Beyer,~B.; Gao,~F.; Scholz,~R.; Poelking,~C.;
  Ortstein,~K.; Günther,~A.~A.; Kasemann,~D.; Andrienko,~D.; Leo,~K. Band
  structure engineering in organic semiconductors. \emph{Science}
  \textbf{2016}, \emph{352}, 1446--1449\relax
\mciteBstWouldAddEndPuncttrue
\mciteSetBstMidEndSepPunct{\mcitedefaultmidpunct}
{\mcitedefaultendpunct}{\mcitedefaultseppunct}\relax
\EndOfBibitem
\bibitem[Zhang \latin{et~al.}(2016)Zhang, Luo, Zhang, Yu, Kuang, Zhang, Meng,
  Luo, Yang, Dong, \latin{et~al.} others]{zhang2016visualizing}
Zhang,~Y.; Luo,~Y.; Zhang,~Y.; Yu,~Y.-J.; Kuang,~Y.-M.; Zhang,~L.; Meng,~Q.-S.;
  Luo,~Y.; Yang,~J.-L.; Dong,~Z.-C., \latin{et~al.}  Visualizing coherent
  intermolecular dipole--dipole coupling in real space. \emph{Nature}
  \textbf{2016}, \emph{531}, 623--627\relax
\mciteBstWouldAddEndPuncttrue
\mciteSetBstMidEndSepPunct{\mcitedefaultmidpunct}
{\mcitedefaultendpunct}{\mcitedefaultseppunct}\relax
\EndOfBibitem
\bibitem[P.Erk(2004)]{P.ErkCrystal}
P.Erk, Experimental Crystal Structure Determination. 2004; CCDC 112723,
  CSD-CUPOCY14\relax
\mciteBstWouldAddEndPuncttrue
\mciteSetBstMidEndSepPunct{\mcitedefaultmidpunct}
{\mcitedefaultendpunct}{\mcitedefaultseppunct}\relax
\EndOfBibitem
\bibitem[Berger \latin{et~al.}(2000)Berger, Fischer, Adolphi, Tierbach, Melev,
  and Schreiber]{berger2000studies}
Berger,~O.; Fischer,~W.-J.; Adolphi,~B.; Tierbach,~S.; Melev,~V.; Schreiber,~J.
  Studies on phase transformations of Cu-phthalocyanine thin films. \emph{J.
  Mater. Sci.: Mater. Electron.} \textbf{2000}, \emph{11}, 331--346\relax
\mciteBstWouldAddEndPuncttrue
\mciteSetBstMidEndSepPunct{\mcitedefaultmidpunct}
{\mcitedefaultendpunct}{\mcitedefaultseppunct}\relax
\EndOfBibitem
\bibitem[de~Oteyza \latin{et~al.}(2006)de~Oteyza, Barrena, Ossó, Sellner, and
  Dosch]{doi:10.1021/ja064641r}
de~Oteyza,~D.~G.; Barrena,~E.; Ossó,~J.~O.; Sellner,~S.; Dosch,~H.
  Thickness-Dependent Structural Transitions in Fluorinated
  Copper-phthalocyanine (F16CuPc) Films. \emph{J. Am. Chem. Soc.}
  \textbf{2006}, \emph{128}, 15052--15053, PMID: 17117832\relax
\mciteBstWouldAddEndPuncttrue
\mciteSetBstMidEndSepPunct{\mcitedefaultmidpunct}
{\mcitedefaultendpunct}{\mcitedefaultseppunct}\relax
\EndOfBibitem
\bibitem[Brown(1968)]{brown1968crystal}
Brown,~C. Crystal structure of $\beta$-copper phthalocyanine. \emph{J. Chem.
  Soc. A} \textbf{1968}, 2488--2493\relax
\mciteBstWouldAddEndPuncttrue
\mciteSetBstMidEndSepPunct{\mcitedefaultmidpunct}
{\mcitedefaultendpunct}{\mcitedefaultseppunct}\relax
\EndOfBibitem
\bibitem[Chai and Head-Gordon(2008)Chai, and Head-Gordon]{chai2008long}
Chai,~J.-D.; Head-Gordon,~M. Long-range corrected hybrid density functionals
  with damped atom--atom dispersion corrections. \emph{Phys. Chem. Chem. Phys.}
  \textbf{2008}, \emph{10}, 6615--6620\relax
\mciteBstWouldAddEndPuncttrue
\mciteSetBstMidEndSepPunct{\mcitedefaultmidpunct}
{\mcitedefaultendpunct}{\mcitedefaultseppunct}\relax
\EndOfBibitem
\bibitem[Frisch \latin{et~al.}(2016)Frisch, Trucks, Schlegel, Scuseria, Robb,
  Cheeseman, Scalmani, Barone, Petersson, Nakatsuji, Li, Caricato, Marenich,
  Bloino, Janesko, Gomperts, Mennucci, Hratchian, Ortiz, Izmaylov, Sonnenberg,
  Williams-Young, Ding, Lipparini, Egidi, Goings, Peng, Petrone, Henderson,
  Ranasinghe, Zakrzewski, Gao, Rega, Zheng, Liang, Hada, Ehara, Toyota, Fukuda,
  Hasegawa, Ishida, Nakajima, Honda, Kitao, Nakai, Vreven, Throssell,
  Montgomery, Peralta, Ogliaro, Bearpark, Heyd, Brothers, Kudin, Staroverov,
  Keith, Kobayashi, Normand, Raghavachari, Rendell, Burant, Iyengar, Tomasi,
  Cossi, Millam, Klene, Adamo, Cammi, Ochterski, Martin, Morokuma, Farkas,
  Foresman, and Fox]{g16}
Frisch,~M.~J. \latin{et~al.}  Gaussian˜16 {R}evision {C}.01. 2016; Gaussian
  Inc. Wallingford CT\relax
\mciteBstWouldAddEndPuncttrue
\mciteSetBstMidEndSepPunct{\mcitedefaultmidpunct}
{\mcitedefaultendpunct}{\mcitedefaultseppunct}\relax
\EndOfBibitem
\bibitem[Lisinetskaya \latin{et~al.}(2016)Lisinetskaya, R{\"o}hr, and
  Mitri{\'c}]{lisinetskaya2016first}
Lisinetskaya,~P.~G.; R{\"o}hr,~M.~I.; Mitri{\'c},~R. First-principles
  simulation of light propagation and exciton dynamics in metal cluster
  nanostructures. \emph{Appl. Phys. B} \textbf{2016}, \emph{122}, 1--12\relax
\mciteBstWouldAddEndPuncttrue
\mciteSetBstMidEndSepPunct{\mcitedefaultmidpunct}
{\mcitedefaultendpunct}{\mcitedefaultseppunct}\relax
\EndOfBibitem
\bibitem[R{\"o}hr \latin{et~al.}(2016)R{\"o}hr, Lisinetskaya, and
  Mitric]{roehr2016excitonic}
R{\"o}hr,~M.~I.; Lisinetskaya,~P.~G.; Mitric,~R. Excitonic properties of
  ordered metal nanocluster arrays: 2D silver clusters at multiporphyrin
  templates. \emph{J. Phys. Chem A} \textbf{2016}, \emph{120}, 4465--4472\relax
\mciteBstWouldAddEndPuncttrue
\mciteSetBstMidEndSepPunct{\mcitedefaultmidpunct}
{\mcitedefaultendpunct}{\mcitedefaultseppunct}\relax
\EndOfBibitem
\bibitem[Lisinetskaya and Mitric(2019)Lisinetskaya, and
  Mitric]{lisinetskaya2019collective}
Lisinetskaya,~P.~G.; Mitric,~R. Collective response in DNA-stabilized silver
  cluster assemblies from first-principles simulations. \emph{J. Phys. Chem.
  Lett.} \textbf{2019}, \emph{10}, 7884--7889\relax
\mciteBstWouldAddEndPuncttrue
\mciteSetBstMidEndSepPunct{\mcitedefaultmidpunct}
{\mcitedefaultendpunct}{\mcitedefaultseppunct}\relax
\EndOfBibitem
\bibitem[Madjet \latin{et~al.}(2006)Madjet, Abdurahman, and
  Renger]{madjet2006intermolecular}
Madjet,~M.; Abdurahman,~A.; Renger,~T. Intermolecular Coulomb couplings from ab
  initio electrostatic potentials: application to optical transitions of
  strongly coupled pigments in photosynthetic antennae and reaction centers.
  \emph{J. Phys. Chem. B} \textbf{2006}, \emph{110}, 17268--17281\relax
\mciteBstWouldAddEndPuncttrue
\mciteSetBstMidEndSepPunct{\mcitedefaultmidpunct}
{\mcitedefaultendpunct}{\mcitedefaultseppunct}\relax
\EndOfBibitem
\bibitem[Jaynes and Cummings(1963)Jaynes, and Cummings]{1443594}
Jaynes,~E.; Cummings,~F. Comparison of quantum and semiclassical radiation
  theories with application to the beam maser. \emph{Proc. IEEE} \textbf{1963},
  \emph{51}, 89--109\relax
\mciteBstWouldAddEndPuncttrue
\mciteSetBstMidEndSepPunct{\mcitedefaultmidpunct}
{\mcitedefaultendpunct}{\mcitedefaultseppunct}\relax
\EndOfBibitem
\bibitem[Gonz\'alez-Tudela \latin{et~al.}(2013)Gonz\'alez-Tudela, Huidobro,
  Mart\'{\i}n-Moreno, Tejedor, and Garc\'{\i}a-Vidal]{PhysRevLett.110.126801}
Gonz\'alez-Tudela,~A.; Huidobro,~P.~A.; Mart\'{\i}n-Moreno,~L.; Tejedor,~C.;
  Garc\'{\i}a-Vidal,~F.~J. Theory of Strong Coupling between Quantum Emitters
  and Propagating Surface Plasmons. \emph{Phys. Rev. Lett.} \textbf{2013},
  \emph{110}, 126801\relax
\mciteBstWouldAddEndPuncttrue
\mciteSetBstMidEndSepPunct{\mcitedefaultmidpunct}
{\mcitedefaultendpunct}{\mcitedefaultseppunct}\relax
\EndOfBibitem
\bibitem[Yuen-Zhou \latin{et~al.}(2016)Yuen-Zhou, Saikin, Zhu, Onbasli, Ross,
  Bulovic, and Baldo]{yuen2016plexciton}
Yuen-Zhou,~J.; Saikin,~S.~K.; Zhu,~T.; Onbasli,~M.~C.; Ross,~C.~A.;
  Bulovic,~V.; Baldo,~M.~A. Plexciton Dirac points and topological modes.
  \emph{Nat. Commun.} \textbf{2016}, \emph{7}, 1--7\relax
\mciteBstWouldAddEndPuncttrue
\mciteSetBstMidEndSepPunct{\mcitedefaultmidpunct}
{\mcitedefaultendpunct}{\mcitedefaultseppunct}\relax
\EndOfBibitem
\bibitem[Yuen-Zhou \latin{et~al.}(2018)Yuen-Zhou, Saikin, and
  Menon]{yuen2018molecular}
Yuen-Zhou,~J.; Saikin,~S.~K.; Menon,~V.~M. Molecular emission near metal
  interfaces: The polaritonic regime. \emph{J. Phys. Chem. Lett.}
  \textbf{2018}, \emph{9}, 6511--6516\relax
\mciteBstWouldAddEndPuncttrue
\mciteSetBstMidEndSepPunct{\mcitedefaultmidpunct}
{\mcitedefaultendpunct}{\mcitedefaultseppunct}\relax
\EndOfBibitem
\bibitem[El-Nahass \latin{et~al.}(2004)El-Nahass, Zeyada, Aziz, and
  El-Ghamaz]{ELNAHASS2004491}
El-Nahass,~M.; Zeyada,~H.; Aziz,~M.; El-Ghamaz,~N. Structural and optical
  properties of thermally evaporated zinc phthalocyanine thin films. \emph{Opt.
  Mater.} \textbf{2004}, \emph{27}, 491--498\relax
\mciteBstWouldAddEndPuncttrue
\mciteSetBstMidEndSepPunct{\mcitedefaultmidpunct}
{\mcitedefaultendpunct}{\mcitedefaultseppunct}\relax
\EndOfBibitem
\bibitem[Wojdyła \latin{et~al.}(2006)Wojdyła, Derkowska, Łukasiak, and
  Bała]{WOJDYLA20063441}
Wojdyła,~M.; Derkowska,~B.; Łukasiak,~Z.; Bała,~W. Absorption and
  photoreflectance spectroscopy of zinc phthalocyanine (ZnPc) thin films grown
  by thermal evaporation. \emph{Mater. Lett.} \textbf{2006}, \emph{60},
  3441--3446\relax
\mciteBstWouldAddEndPuncttrue
\mciteSetBstMidEndSepPunct{\mcitedefaultmidpunct}
{\mcitedefaultendpunct}{\mcitedefaultseppunct}\relax
\EndOfBibitem
\bibitem[Hosokai \latin{et~al.}(2010)Hosokai, Gerlach, Hinderhofer, Frank,
  Ligorio, Heinemeyer, Vorobiev, and Schreiber]{hosokai2010simultaneous}
Hosokai,~T.; Gerlach,~A.; Hinderhofer,~A.; Frank,~C.; Ligorio,~G.;
  Heinemeyer,~U.; Vorobiev,~A.; Schreiber,~F. Simultaneous in situ measurements
  of x-ray reflectivity and optical spectroscopy during organic semiconductor
  thin film growth. \emph{Appl. Phys. Lett.} \textbf{2010}, \emph{97},
  170\relax
\mciteBstWouldAddEndPuncttrue
\mciteSetBstMidEndSepPunct{\mcitedefaultmidpunct}
{\mcitedefaultendpunct}{\mcitedefaultseppunct}\relax
\EndOfBibitem
\bibitem[Pandey \latin{et~al.}(2012)Pandey, Rochford, Keeble, Rourke, Jones,
  Beanland, and Wilson]{pandey2012resolving}
Pandey,~P.~A.; Rochford,~L.~A.; Keeble,~D.~S.; Rourke,~J.~P.; Jones,~T.~S.;
  Beanland,~R.; Wilson,~N.~R. Resolving the nanoscale morphology and
  crystallographic structure of molecular thin films: F16CuPc on graphene
  oxide. \emph{Chem. Mater.} \textbf{2012}, \emph{24}, 1365--1370, CCDC 853132,
  CSD-FUJPUW01\relax
\mciteBstWouldAddEndPuncttrue
\mciteSetBstMidEndSepPunct{\mcitedefaultmidpunct}
{\mcitedefaultendpunct}{\mcitedefaultseppunct}\relax
\EndOfBibitem
\bibitem[Jiang \latin{et~al.}(2017)Jiang, Hu, Ye, Li, Li, Zhang, Li, Dong, Hu,
  and Kloc]{jiang2017molecular}
Jiang,~H.; Hu,~P.; Ye,~J.; Li,~Y.; Li,~H.; Zhang,~X.; Li,~R.; Dong,~H.; Hu,~W.;
  Kloc,~C. Molecular Crystal Engineering: Tuning Organic Semiconductor from
  p-type to n-type by Adjusting Their Substitutional Symmetry. \emph{Adv.
  Mater.} \textbf{2017}, \emph{29}, 1605053, CCDC 1482757, CSD-SOTJAP; CCDC
  1511393, CSD-SOTJUJ\relax
\mciteBstWouldAddEndPuncttrue
\mciteSetBstMidEndSepPunct{\mcitedefaultmidpunct}
{\mcitedefaultendpunct}{\mcitedefaultseppunct}\relax
\EndOfBibitem
\bibitem[Savolainen \latin{et~al.}(2008)Savolainen, van~der Linden, Dijkhuizen,
  and Herek]{savolainen2008characterizing}
Savolainen,~J.; van~der Linden,~D.; Dijkhuizen,~N.; Herek,~J.~L. Characterizing
  the functional dynamics of zinc phthalocyanine from femtoseconds to
  nanoseconds. \emph{J. Photochem. Photobiol. A} \textbf{2008}, \emph{196},
  99--105\relax
\mciteBstWouldAddEndPuncttrue
\mciteSetBstMidEndSepPunct{\mcitedefaultmidpunct}
{\mcitedefaultendpunct}{\mcitedefaultseppunct}\relax
\EndOfBibitem
\bibitem[Yanagisawa \latin{et~al.}(2013)Yanagisawa, Yasuda, Inagaki, Morikawa,
  Manseki, and Yanagida]{doi:10.1021/jp407608w}
Yanagisawa,~S.; Yasuda,~T.; Inagaki,~K.; Morikawa,~Y.; Manseki,~K.;
  Yanagida,~S. Intermolecular Interaction as the Origin of Red Shifts in
  Absorption Spectra of Zinc-Phthalocyanine from First-Principles. \emph{J.
  Phys. Chem. A} \textbf{2013}, \emph{117}, 11246--11253, PMID: 24106753\relax
\mciteBstWouldAddEndPuncttrue
\mciteSetBstMidEndSepPunct{\mcitedefaultmidpunct}
{\mcitedefaultendpunct}{\mcitedefaultseppunct}\relax
\EndOfBibitem
\bibitem[Peisert \latin{et~al.}(2001)Peisert, Schwieger, Auerhammer, Knupfer,
  Golden, Fink, Bressler, and Mast]{peisert2001order}
Peisert,~H.; Schwieger,~T.; Auerhammer,~J.; Knupfer,~M.; Golden,~M.; Fink,~J.;
  Bressler,~P.; Mast,~M. Order on disorder: Copper phthalocyanine thin films on
  technical substrates. \emph{J. Appl. Phys.} \textbf{2001}, \emph{90},
  466--469\relax
\mciteBstWouldAddEndPuncttrue
\mciteSetBstMidEndSepPunct{\mcitedefaultmidpunct}
{\mcitedefaultendpunct}{\mcitedefaultseppunct}\relax
\EndOfBibitem
\bibitem[Ikame \latin{et~al.}(2005)Ikame, Kanai, Ouchi, Ito, Fujimori, and
  Seki]{IKAME2005373}
Ikame,~T.; Kanai,~K.; Ouchi,~Y.; Ito,~E.; Fujimori,~A.; Seki,~K. Molecular
  orientation of F16ZnPc deposited on Au and Mg substrates studied by NEXAFS
  and IRRAS. \emph{Chem. Phys. Lett.} \textbf{2005}, \emph{413}, 373--378\relax
\mciteBstWouldAddEndPuncttrue
\mciteSetBstMidEndSepPunct{\mcitedefaultmidpunct}
{\mcitedefaultendpunct}{\mcitedefaultseppunct}\relax
\EndOfBibitem
\bibitem[de~Oteyza \latin{et~al.}(2006)de~Oteyza, Barrena, Sellner, Ossó, and
  Dosch]{doi:10.1021/jp061889u}
de~Oteyza,~D.~G.; Barrena,~E.; Sellner,~S.; Ossó,~J.~O.; Dosch,~H. Structural
  Rearrangements During the Initial Growth Stages of Organic Thin Films of
  F16CuPc on SiO2. \emph{J. Phys. Chem. B} \textbf{2006}, \emph{110},
  16618--16623, PMID: 16913797\relax
\mciteBstWouldAddEndPuncttrue
\mciteSetBstMidEndSepPunct{\mcitedefaultmidpunct}
{\mcitedefaultendpunct}{\mcitedefaultseppunct}\relax
\EndOfBibitem
\bibitem[Lozzi and Santucci(2003)Lozzi, and Santucci]{lozzi2003xps}
Lozzi,~L.; Santucci,~S. XPS study of the FCuPc/SiO2 interface. \emph{Surf.
  Sci.} \textbf{2003}, \emph{532}, 976--981\relax
\mciteBstWouldAddEndPuncttrue
\mciteSetBstMidEndSepPunct{\mcitedefaultmidpunct}
{\mcitedefaultendpunct}{\mcitedefaultseppunct}\relax
\EndOfBibitem
\end{mcitethebibliography}


\providecommand{\latin}[1]{#1}
\makeatletter
\providecommand{\doi}
  {\begingroup\let\do\@makeother\dospecials
  \catcode`\{=1 \catcode`\}=2 \doi@aux}
\providecommand{\doi@aux}[1]{\endgroup\texttt{#1}}
\makeatother
\providecommand*\mcitethebibliography{\thebibliography}
\csname @ifundefined\endcsname{endmcitethebibliography}
  {\let\endmcitethebibliography\endthebibliography}{}
\begin{mcitethebibliography}{19}
\providecommand*\natexlab[1]{#1}
\providecommand*\mciteSetBstSublistMode[1]{}
\providecommand*\mciteSetBstMaxWidthForm[2]{}
\providecommand*\mciteBstWouldAddEndPuncttrue
  {\def\EndOfBibitem{\unskip.}}
\providecommand*\mciteBstWouldAddEndPunctfalse
  {\let\EndOfBibitem\relax}
\providecommand*\mciteSetBstMidEndSepPunct[3]{}
\providecommand*\mciteSetBstSublistLabelBeginEnd[3]{}
\providecommand*\EndOfBibitem{}
\mciteSetBstSublistMode{f}
\mciteSetBstMaxWidthForm{subitem}{(\alph{mcitesubitemcount})}
\mciteSetBstSublistLabelBeginEnd
  {\mcitemaxwidthsubitemform\space}
  {\relax}
  {\relax}

\bibitem[Madjet \latin{et~al.}(2006)Madjet, Abdurahman, and Renger]{TranCharge}
Madjet,~M.~E.; Abdurahman,~A.; Renger,~T. Intermolecular Coulomb couplings from
  ab initio electrostatic potentials: Application to optical transitions of
  strongly coupled pigments in photosynthetic antennae and reaction centers.
  \emph{J. Phys. Chem. B} \textbf{2006}, \emph{110}, 17268--17281\relax
\mciteBstWouldAddEndPuncttrue
\mciteSetBstMidEndSepPunct{\mcitedefaultmidpunct}
{\mcitedefaultendpunct}{\mcitedefaultseppunct}\relax
\EndOfBibitem
\bibitem[Lisinetskaya and Mitri\'{c}(2014)Lisinetskaya, and
  Mitri\'{c}]{Polina2014}
Lisinetskaya,~P.~G.; Mitri\'{c},~R. Ab Initio Simulations of Light Propagation
  in Silver Cluster Nanostructures. \emph{Phys. Rev. B} \textbf{2014},
  \emph{89}, 035433\relax
\mciteBstWouldAddEndPuncttrue
\mciteSetBstMidEndSepPunct{\mcitedefaultmidpunct}
{\mcitedefaultendpunct}{\mcitedefaultseppunct}\relax
\EndOfBibitem
\bibitem[Lisinetskaya \latin{et~al.}(2016)Lisinetskaya, R\"{o}hr, and
  Mitri\'{c}]{Polina2016ring}
Lisinetskaya,~P.~G.; R\"{o}hr,~M. I.~S.; Mitri\'{c},~R. {First-Principles
  Simulation of Light Propagation and Exciton Dynamics in Metal Cluster
  Nanostructures}. \emph{Appl. Phys. B} \textbf{2016}, \emph{122}, 175\relax
\mciteBstWouldAddEndPuncttrue
\mciteSetBstMidEndSepPunct{\mcitedefaultmidpunct}
{\mcitedefaultendpunct}{\mcitedefaultseppunct}\relax
\EndOfBibitem
\bibitem[R\"{o}hr \latin{et~al.}(2016)R\"{o}hr, Lisinetskaya, and
  Mitri\'{c}]{Polina2016ag3}
R\"{o}hr,~M. I.~S.; Lisinetskaya,~P.~G.; Mitri\'{c},~R. {Excitonic Properties
  of Ordered Metal Nanocluster Arrays: 2D Silver Clusters at Multiporphyrin
  Templates}. \emph{J. Phys. Chem. A} \textbf{2016}, \emph{120}, 4465--4472,
  doi: 10.1021/acs.jpca.6b04243\relax
\mciteBstWouldAddEndPuncttrue
\mciteSetBstMidEndSepPunct{\mcitedefaultmidpunct}
{\mcitedefaultendpunct}{\mcitedefaultseppunct}\relax
\EndOfBibitem
\bibitem[{Lisinetskaya Polina G.} and {Mitri\'{c} Roland}(2019){Lisinetskaya
  Polina G.}, and {Mitri\'{c} Roland}]{Polina2019}
{Lisinetskaya Polina G.},; {Mitri\'{c} Roland}, {Collective Response in
  DNA-Stabilized Silver Cluster Assemblies from First-Principles Simulations}.
  \emph{J. Phys. Chem. Lett.} \textbf{2019}, \emph{10}, 7884--7889, doi:
  10.1021/acs.jpclett.9b03136\relax
\mciteBstWouldAddEndPuncttrue
\mciteSetBstMidEndSepPunct{\mcitedefaultmidpunct}
{\mcitedefaultendpunct}{\mcitedefaultseppunct}\relax
\EndOfBibitem
\bibitem[Jaynes and Cummings(1963)Jaynes, and Cummings]{1443594}
Jaynes,~E.; Cummings,~F. Comparison of quantum and semiclassical radiation
  theories with application to the beam maser. \emph{Proceedings of the IEEE}
  \textbf{1963}, \emph{51}, 89--109\relax
\mciteBstWouldAddEndPuncttrue
\mciteSetBstMidEndSepPunct{\mcitedefaultmidpunct}
{\mcitedefaultendpunct}{\mcitedefaultseppunct}\relax
\EndOfBibitem
\bibitem[Gonz\'alez-Tudela \latin{et~al.}(2013)Gonz\'alez-Tudela, Huidobro,
  Mart\'{\i}n-Moreno, Tejedor, and Garc\'{\i}a-Vidal]{PhysRevLett.110.126801}
Gonz\'alez-Tudela,~A.; Huidobro,~P.~A.; Mart\'{\i}n-Moreno,~L.; Tejedor,~C.;
  Garc\'{\i}a-Vidal,~F.~J. Theory of Strong Coupling between Quantum Emitters
  and Propagating Surface Plasmons. \emph{Physical review letters}
  \textbf{2013}, \emph{110}, 126801\relax
\mciteBstWouldAddEndPuncttrue
\mciteSetBstMidEndSepPunct{\mcitedefaultmidpunct}
{\mcitedefaultendpunct}{\mcitedefaultseppunct}\relax
\EndOfBibitem
\bibitem[Yuen-Zhou \latin{et~al.}(2016)Yuen-Zhou, Saikin, Zhu, Onbasli, Ross,
  Bulovic, and Baldo]{yuen2016plexciton}
Yuen-Zhou,~J.; Saikin,~S.~K.; Zhu,~T.; Onbasli,~M.~C.; Ross,~C.~A.;
  Bulovic,~V.; Baldo,~M.~A. Plexciton Dirac points and topological modes.
  \emph{Nature communications} \textbf{2016}, \emph{7}, 1--7\relax
\mciteBstWouldAddEndPuncttrue
\mciteSetBstMidEndSepPunct{\mcitedefaultmidpunct}
{\mcitedefaultendpunct}{\mcitedefaultseppunct}\relax
\EndOfBibitem
\bibitem[Yuen-Zhou \latin{et~al.}(2018)Yuen-Zhou, Saikin, and
  Menon]{yuen2018molecular}
Yuen-Zhou,~J.; Saikin,~S.~K.; Menon,~V.~M. Molecular emission near metal
  interfaces: The polaritonic regime. \emph{The journal of physical chemistry
  letters} \textbf{2018}, \emph{9}, 6511--6516\relax
\mciteBstWouldAddEndPuncttrue
\mciteSetBstMidEndSepPunct{\mcitedefaultmidpunct}
{\mcitedefaultendpunct}{\mcitedefaultseppunct}\relax
\EndOfBibitem
\bibitem[Bellessa \latin{et~al.}(2004)Bellessa, Bonnand, Plenet, and
  Mugnier]{PhysRevLett.93.036404}
Bellessa,~J.; Bonnand,~C.; Plenet,~J.~C.; Mugnier,~J. Strong Coupling between
  Surface Plasmons and Excitons in an Organic Semiconductor. \emph{Physical
  review letters} \textbf{2004}, \emph{93}, 036404\relax
\mciteBstWouldAddEndPuncttrue
\mciteSetBstMidEndSepPunct{\mcitedefaultmidpunct}
{\mcitedefaultendpunct}{\mcitedefaultseppunct}\relax
\EndOfBibitem
\bibitem[Gonz\'alez-Tudela \latin{et~al.}(2013)Gonz\'alez-Tudela, Huidobro,
  Mart\'{\i}n-Moreno, Tejedor, and Garc\'{\i}a-Vidal]{GonzalesTudela2013}
Gonz\'alez-Tudela,~A.; Huidobro,~P.~A.; Mart\'{\i}n-Moreno,~L.; Tejedor,~C.;
  Garc\'{\i}a-Vidal,~F.~J. Theory of Strong Coupling between Quantum Emitters
  and Propagating Surface Plasmons. \emph{Phys. Rev. Lett.} \textbf{2013},
  \emph{110}, 126801\relax
\mciteBstWouldAddEndPuncttrue
\mciteSetBstMidEndSepPunct{\mcitedefaultmidpunct}
{\mcitedefaultendpunct}{\mcitedefaultseppunct}\relax
\EndOfBibitem
\bibitem[{Yuen-Zhou Joel} \latin{et~al.}(2018){Yuen-Zhou Joel}, {Saikin Semion
  K.}, and {Menon Vinod M.}]{YuenZhou2018}
{Yuen-Zhou Joel},; {Saikin Semion K.},; {Menon Vinod M.}, {Molecular Emission
  near Metal Interfaces: The Polaritonic Regime}. \emph{J Phys. Chem. Lett.}
  \textbf{2018}, \emph{9}, 6511--6516, doi: 10.1021/acs.jpclett.8b02980\relax
\mciteBstWouldAddEndPuncttrue
\mciteSetBstMidEndSepPunct{\mcitedefaultmidpunct}
{\mcitedefaultendpunct}{\mcitedefaultseppunct}\relax
\EndOfBibitem
\bibitem[Chai and Head-Gordon(2008)Chai, and Head-Gordon]{Chai2008}
Chai,~J.-D.; Head-Gordon,~M. {Long-range corrected hybrid density functionals
  with damped atom-atom dispersion corrections}. \emph{Phys. Chem. Chem. Phys.}
  \textbf{2008}, \emph{10}, 6615--6620\relax
\mciteBstWouldAddEndPuncttrue
\mciteSetBstMidEndSepPunct{\mcitedefaultmidpunct}
{\mcitedefaultendpunct}{\mcitedefaultseppunct}\relax
\EndOfBibitem
\bibitem[Frisch \latin{et~al.}(2016)Frisch, Trucks, Schlegel, Scuseria, Robb,
  Cheeseman, Scalmani, Barone, Petersson, Nakatsuji, Li, Caricato, Marenich,
  Bloino, Janesko, Gomperts, Mennucci, Hratchian, Ortiz, Izmaylov, Sonnenberg,
  Williams-Young, Ding, Lipparini, Egidi, Goings, Peng, Petrone, Henderson,
  Ranasinghe, Zakrzewski, Gao, Rega, Zheng, Liang, Hada, Ehara, Toyota, Fukuda,
  Hasegawa, Ishida, Nakajima, Honda, Kitao, Nakai, Vreven, Throssell,
  Montgomery, Peralta, Ogliaro, Bearpark, Heyd, Brothers, Kudin, Staroverov,
  Keith, Kobayashi, Normand, Raghavachari, Rendell, Burant, Iyengar, Tomasi,
  Cossi, Millam, Klene, Adamo, Cammi, Ochterski, Martin, Morokuma, Farkas,
  Foresman, and Fox]{g16}
Frisch,~M.~J. \latin{et~al.}  Gaussian16 {R}evision {C}.01. 2016; Gaussian Inc.
  Wallingford CT\relax
\mciteBstWouldAddEndPuncttrue
\mciteSetBstMidEndSepPunct{\mcitedefaultmidpunct}
{\mcitedefaultendpunct}{\mcitedefaultseppunct}\relax
\EndOfBibitem
\bibitem[Ngoc \latin{et~al.}(2015)Ngoc, Wiedemair, van~den Berg, and
  Carlen]{Ngoc2015}
Ngoc,~L. L.~T.; Wiedemair,~J.; van~den Berg,~A.; Carlen,~E.~T.
  {Plasmon-modulated photoluminescence from gold nanostructures and its
  dependence on plasmon resonance, excitation energy, and band structure}.
  \emph{Opt. Express} \textbf{2015}, \emph{23}, 5547--5564\relax
\mciteBstWouldAddEndPuncttrue
\mciteSetBstMidEndSepPunct{\mcitedefaultmidpunct}
{\mcitedefaultendpunct}{\mcitedefaultseppunct}\relax
\EndOfBibitem
\bibitem[P.Erk(2004)]{CuPc}
P.Erk, CCDC 112723: Experimental Crystal Structure Determination. 2004; DOI:
  10.5517/cc3s97d\relax
\mciteBstWouldAddEndPuncttrue
\mciteSetBstMidEndSepPunct{\mcitedefaultmidpunct}
{\mcitedefaultendpunct}{\mcitedefaultseppunct}\relax
\EndOfBibitem
\bibitem[Jiang \latin{et~al.}(2017)Jiang, Hu, Ye, Li, Li, Zhang, Li, Dong, Hu,
  and Kloc]{F4F8CuPc}
Jiang,~H.; Hu,~P.; Ye,~J.; Li,~Y.; Li,~H.; Zhang,~X.; Li,~R.; Dong,~H.; Hu,~W.;
  Kloc,~C. {Molecular Crystal Engineering: Tuning Organic Semiconductor from
  p-type to n-type by Adjusting Their Substitutional Symmetry}. \emph{Advanced
  Materials} \textbf{2017}, \emph{29}, 1605053\relax
\mciteBstWouldAddEndPuncttrue
\mciteSetBstMidEndSepPunct{\mcitedefaultmidpunct}
{\mcitedefaultendpunct}{\mcitedefaultseppunct}\relax
\EndOfBibitem
\bibitem[{Pandey Priyanka A.} \latin{et~al.}(2012){Pandey Priyanka A.},
  {Rochford Luke A.}, {Keeble Dean S.}, {Rourke Jonathan P.}, {Jones Tim S.},
  {Beanland Richard}, and {Wilson Neil R.}]{F16CuPc}
{Pandey Priyanka A.},; {Rochford Luke A.},; {Keeble Dean S.},; {Rourke Jonathan
  P.},; {Jones Tim S.},; {Beanland Richard},; {Wilson Neil R.}, {Resolving the
  Nanoscale Morphology and Crystallographic Structure of Molecular Thin Films:
  F16CuPc on Graphene Oxide}. \emph{Chemistry of Materials} \textbf{2012},
  \emph{24}, 1365--1370, doi: 10.1021/cm300073v\relax
\mciteBstWouldAddEndPuncttrue
\mciteSetBstMidEndSepPunct{\mcitedefaultmidpunct}
{\mcitedefaultendpunct}{\mcitedefaultseppunct}\relax
\EndOfBibitem
\end{mcitethebibliography}

\end{document}

\part*{Supporting Information}

\section{Theoretical methods}

\subsection{Construction of excitonic Hamiltonian}

\renewcommand{\thefigure}{S\arabic{figure}}
\renewcommand{\thetable}{S\arabic{table}}
\renewcommand{\theequation}{S\arabic{equation}}

The excitonic Hamiltonian of an aggregate consisting of $N$ molecules
with particular spatial organization can be constructed in the following
way: 

\begin{equation}
\mathrm{H}_{Exc}=\underset{{\scriptscriptstyle I=1}}{\overset{{\scriptscriptstyle N}}{\sum}}\mathrm{H}_{I}^{0}+\underset{{\scriptscriptstyle I=1}}{\overset{{\scriptscriptstyle N}}{\sum}}\underset{{\scriptscriptstyle J>I}}{\sum}\mathrm{V}_{IJ},\label{eq:total Hamiltonian}
\end{equation}
were $\mathrm{H}_{I}^{0}$ is the electronic Hamiltonian of the $I$-th
molecule and $\mathrm{V}_{IJ}$ is the operator describing the pair-wise
interactions between the molecules. Within the current model the interaction
is considered as pure electromagnetic and the latter operator thus
has the Coulomb form (in atomic units): 

\begin{equation}
{\textstyle \mathrm{V}}_{IJ}=\underset{{\scriptstyle {\scriptscriptstyle ab}}}{\sum}\frac{1}{\left|\mathbf{r}_{a}-\mathbf{r}_{b}\right|}-\underset{{\scriptscriptstyle aB}}{\sum}\frac{Z_{B}}{\left|\mathbf{r}_{a}-\mathbf{r}_{B}\right|}-\underset{{\scriptscriptstyle Ab}}{\sum}\frac{Z_{A}}{\left|\mathbf{r}_{A}-\mathbf{r}_{b}\right|}+\underset{{\scriptscriptstyle AB}}{\sum}\frac{Z_{A}Z_{B}}{\left|\mathbf{r}_{A}-\mathbf{r}_{B}\right|}.\label{eq:interaction operator}
\end{equation}
Here, $\mathbf{r}$ and $Z$ denote the position (the bold font stands
for a vector) and charge of electrons and nuclei, the indices $a$,
$b$ ($A$, $B$) run over all electrons (nuclei) of the molecules
$I$ and $J$, respectively. The natural basis to represent the Hamiltonian
matrix is the product-state basis built up from the eigenfunctions
of the individual molecules:

\begin{equation}
\left|\phi_{i\dots j\ldots z}\right\rangle =\left|\Psi_{i}^{1}\right\rangle \otimes\ldots\otimes\left|\Psi_{j}^{I}\right\rangle \otimes\ldots\otimes\left|\Psi_{z}^{N}\right\rangle ,\label{eq:basis}
\end{equation}
where the indices $i\dots j\dots z$ run over all electronic states
of each individual molecule and the superscripts 1, 2, ..., $N$ denote
the index number of the molecule in the aggregate. The matrix element
of the interaction operator can be then represented

\begin{multline}
\left\langle \phi_{i\dots j\ldots z}\right|{\textstyle \mathrm{V}}_{IJ}\left|\phi_{i'\dots j'\ldots z'}\right\rangle =\left[\int\mathrm{d}\mathbf{r}_{a}\mathrm{d}\mathbf{r}_{b}\frac{\rho_{ii'}^{I}\left(\mathbf{r}_{a}\right)\rho_{jj'}^{J}\left(\mathbf{r}_{b}\right)}{\left|\mathbf{r}_{a}-\mathbf{r}_{b}\right|}-\underset{{\scriptscriptstyle B}}{\sum}\int\mathrm{d}\mathbf{r}_{a}\frac{\rho_{ii'}^{I}\left(\mathbf{r}_{a}\right)Z_{B}}{\left|\mathbf{r}_{a}-\mathbf{r}_{B}\right|}\delta_{jj'}-\right.\\
\left.\underset{{\scriptscriptstyle A}}{\sum}\int\mathrm{d}\mathbf{r}_{b}\frac{\rho_{jj'}^{I}\left(\mathbf{r}_{b}\right)Z_{A}}{\left|\mathbf{r}_{A}-\mathbf{r}_{b}\right|}\delta_{jj'}+\underset{{\scriptscriptstyle AB}}{\sum}\frac{Z_{A}Z_{B}}{\left|\mathbf{r}_{A}-\mathbf{r}_{B}\right|}\right]\delta_{kk'}\ldots\delta_{zz'}.\label{eq:interaction ME}
\end{multline}

In derivation of Eq. (\ref{eq:interaction ME}) we have used the orthogonality
and the antisymmetricity of the single-molecule electron wave-functions
$\left|\Psi_{i}^{I}\right\rangle $ and introduced the one-electron
($i=i'$) or transition ($i\neq i'$) density:

\begin{equation}
\rho_{ii'}^{I}\left(\mathbf{r}_{1}\right)=N_{I}\dotsint\Psi_{i}^{I*}\left(\mathbf{x}\right)\Psi_{i'}^{I}\left(\mathbf{x}\right)\mathrm{d}\mathbf{x}_{2}\dots\mathrm{d}\mathbf{x}_{N_{I}}\mathrm{d}\sigma_{1},\label{eq:TD integral-1}
\end{equation}
where $\mathbf{x}$ stands for spatial and spin coordinates of all
electrons in the cluster, $N_{I}$ is the number of electrons, the
integration runs over all spatial and spin coordinates of all electrons
except the spatial coordinates of the first one. The calculation of
the matrix element (\ref{eq:interaction ME}) is very computationally
demanding for large molecules, therefore the transition charge approximation
\cite{TranCharge} has been employed to calculate the interaction
elements. Within this approximation, partial charges are obtained
by fitting the electrostatic potential produced by one-electron and
transition densities of a molecule with a set of point charges located
at the positions of the nuclei. In our approach we construct the full
matrix of transition charges $q_{ii'}^{I,A}$ between all electronic
states of individual molecule using the following fit:

\begin{equation}
\varphi{}_{ii'}^{I}\left(\mathbf{r}\right)=-\int\mathrm{d}\mathbf{r}'\frac{\rho_{ii'}^{I}\left(\mathbf{r}'\right)}{\left|\mathbf{r}-\mathbf{r}'\right|}+\underset{{\scriptscriptstyle A}}{\sum}\frac{Z_{A}}{\left|\mathbf{r}-\mathbf{r}_{A}\right|}\delta_{ii'}\approx\underset{{\scriptscriptstyle A}}{\sum}\frac{q_{ii'}^{I,A}}{\left|\mathbf{r}-\mathbf{r}_{A}\right|}.\label{eq:trans charges}
\end{equation}

This allows to dramatically simplify the interaction operator (\ref{eq:interaction operator}),
which now reads

\begin{equation}
\mathrm{V}_{IJ}=\underset{{\scriptscriptstyle AB}}{\sum}\underset{{\scriptscriptstyle ii'}}{\sum}\underset{{\scriptscriptstyle jj'}}{\sum}\frac{\left|\Psi_{i}^{I}\right\rangle \otimes\left|\Psi_{j}^{J}\right\rangle q_{ii'}^{I,A}q_{jj'}^{J,B}\left\langle \Psi_{i'}^{I}\right|\otimes\left\langle \Psi_{j'}^{J}\right|}{\left|\mathbf{r}_{A}-\mathbf{r}_{B}\right|}.\label{eq:inter operator}
\end{equation}
This approximation is superior to the dipole-dipole approximation,
since it aims to mimic the spatial distribution of the electron density
in a molecule and thus accounts for higher multipole moments as well.

Necessary prerequisites needed to construct the matrix representation
of the Hamiltonian (\ref{eq:total Hamiltonian}) are the electronic
state energies and wave-functions of individual molecules which constitute
the aggregate. Obtaining of these quantities in the frame of linear-response
TDDFT is described in details elsewhere \cite{Polina2014,Polina2016ring,Polina2016ag3,Polina2019}.

\subsection{Stationary absorption spectra}

The direct diagonalization of the Hamiltonian (\ref{eq:total Hamiltonian})
provides excitonic eigenstates $E_{\alpha}$ and eigenfunctions $\left|\psi_{\alpha}\right\rangle $
of the molecular aggregate,

\begin{equation}
\left|\psi_{\alpha}\right\rangle =\underset{{\scriptscriptstyle i\dots j\dots z}}{\sum}C_{i\dots j\dots z}^{\alpha}\left|\Psi_{i}^{1}\right\rangle \otimes\ldots\otimes\left|\Psi_{j}^{I}\right\rangle \otimes\ldots\otimes\left|\Psi_{z}^{N}\right\rangle ,\label{eq:aggregate wavefunction}
\end{equation}
which allows to simulate it's absorption spectrum. The excitation
energies are equal to the energy differences between the ground and
excited states $\hbar\omega_{\alpha}=E_{\alpha}-E_{0}$, and the equation
for collective excitonic transition dipole moments $\mathbf{M}_{\beta\alpha}$ needed to calculate the oscillator strengths of the corresponding transitions
$f_{\alpha}=\frac{2}{3}\omega_{\alpha}\left|\mathbf{M}_{0\alpha}\right|^{2}$ can be derived using Eq. (\ref{eq:aggregate wavefunction})

\begin{equation}
\mathbf{M}_{\beta\alpha}=\left\langle \psi_{\beta}\right|\underset{{\scriptstyle I}}{\sum}\boldsymbol{\mu}_{I}\left|\psi_{\alpha}\right\rangle =\underset{{\scriptstyle {\scriptscriptstyle I}}}{\sum}\underset{{\scriptstyle i\ldots j\ldots z}}{\sum}\underset{{\scriptstyle j'}}{\sum}C_{i\dots j\dots z}^{*\beta}C_{i\dots j'\dots z}^{\alpha}\boldsymbol{\mu}_{jj'}^{I}.\label{eq:TDM}
\end{equation}
Here, the full matrix of transition and stationary dipole moments
of each individual molecule $\boldsymbol{\mu}_{jj'}^{I}$ can be obtained
from the corresponding set of transition charges 
\begin{equation}
\boldsymbol{\mu}_{jj'}^{I}=\underset{{\scriptscriptstyle A}}{\sum}q_{jj'}^{I,A}\mathbf{r}_{A}.\label{eq:TDM from TC}
\end{equation}

\subsection{Interaction with plasmonic field}

\subsubsection{Disordered molecular emitters }

The strong coupling phenomena, such as interaction of a surface plasmon
polariton (SPP) with a set of molecular emitters, are usually described
by the Jaynes-Cumming model \cite{1443594}. The model can
be adopted to the situation of spatially delocalized SPP coupled to
a localized exciton \cite{PhysRevLett.110.126801, yuen2016plexciton, yuen2018molecular}. The related Hamilton operator
of the system $\mathrm{H}_{\mathbf{k}}^{N}$ contains the contribution
of all $N$ emitters, 

\begin{equation}
\mathrm{H}_{\mathbf{k}}^{N}=\mathrm{H}_{SPP}+\mathrm{H}_{Exc}+\mathrm{H}_{C}.\label{eq:Plexciton hamiltonian}
\end{equation}
Here, $\mathrm{H}_{Exc}=E_{Exc}D{}_{\mathbf{k}}^{\dagger}D_{\mathbf{k}}$
represents the excitonic part with the creation and annihilation operators
$D_{\mathbf{k}}^{\dagger}$ and $D_{\mathbf{k}}$, respectively, and
the related excitation energy $E_{Exc}$. The 2D surface plasmon polariton
and its energy dispersion $E_{SPP}=\hbar\omega\left(\mathbf{k}\right)$
are accounted for by $\mathrm{H}_{SPP}=\hbar\omega\left(\mathbf{k}\right)a{}_{\mathbf{k}}^{\dagger}a_{\mathbf{k}}$
with $a_{\mathbf{k}}^{\dagger}$ and $a_{\mathbf{k}}$ being the creation
and annihilation operators of the collective plasmonic mode at wavevector
$\mathbf{k}$. The coupling of excitonic and plasmonic states is described
by $\mathrm{H}_{C}$ and can be interpreted as interaction between
of the SPP and the local dipole moment of the molecular emitters 

\begin{equation}
H_{C}=g_{\boldsymbol{\mu}}^{N}\left(\mathbf{k}\right)\left[a_{\mathbf{k}}D_{\mathbf{k}}^{\dagger}+a_{\mathbf{k}}^{\dagger}D_{\mathbf{k}}\right].\label{eq: plexciton interaction}
\end{equation}
The parameter $g_{\boldsymbol{\mu}}^{N}\left(\mathbf{k}\right)$ contains
the contributions of all $N$ quantum emitters of dipole moment $\boldsymbol{\mu}$
to the coupling. However, to render this problem mathematically accessible,
this discrete picture has to be translated into a volume density of
quantum emitters, i.e. their continuous distribution within a layer
of given thickness $d=(z_{0}+d)-z_{0}$ with $z_{0}$ defining the
absolute position of the hybrid bilayer interface. This approach provides
an expression for the effective coupling constant

\begin{equation}
V=g_{\boldsymbol{\mu}}^{N}\left(\mathbf{k}\right)=\sqrt{n\stackrel[{\scriptscriptstyle z_{0}}]{{\scriptscriptstyle z_{0}+d}}{\varint}\left|g_{\boldsymbol{\mu}}\left(\mathbf{k};z\right)\right|^{2}\mathrm{d}z}.\label{eq:Coupling constant}
\end{equation}
Accordingly, the coupling constant depends on the volume density of
the emitters 

\begin{equation}
n=\frac{N_{S}N_{L}}{Ad}\label{eq: emitter density}
\end{equation}
with $A$ being the projected area of the emissive layer on the metal-dielectric
interface. The number of layers is indicated by $N_{L}$ and the related
number of quantum emitters per layer by $N_{S}$. In particular, 

\begin{equation}
g_{\boldsymbol{\mu}}\left(\mathbf{k};z\right)=\sqrt{\frac{\omega\left(\mathbf{k}\right)}{2\epsilon_{0}L\left(\mathbf{k}\right)}}e^{-k_{z}z}\left(\mathbf{e}_{\mathbf{k}}+i\frac{\left|\mathbf{k}\right|}{k_{z}}\mathbf{e}_{z}\right)\cdot\boldsymbol{\mu}\label{eq:coupling g}
\end{equation}
describes the coupling between a molecular transition dipole $\boldsymbol{\mu}$
located at distance $z$ from the metal-dielectric interface and the
electric field of an SPP, $\mathbf{e}_{\mathbf{k}}$ and $\mathbf{e}_{z}$
denote the unit vectors in the direction of the wavevector $\mathbf{k}$
(in-plane) and perpendicular to the metal-dielectric interface (out-of-plane),
$k_{z}=\sqrt{\left|\mathbf{k}\right|^{2}-\epsilon_{d}\omega^{2}\left(\mathbf{k}\right)c^{-2}}$
is the vertical component of the wave vector $\mathbf{k}$. Evidently,
the z-dependence of the coupling is strongly influenced by the exponential
decay $e^{-k_{z}z}$ of the evanescent electromagnetic SPP component
along the interface normal. In Eq. (\ref{eq:Coupling constant}),
$L\left(\mathbf{k}\right)$ denotes the effective in-plane length
of the plasmon mode. 

Matrix representation of the Hamiltonian (\ref{eq:Plexciton hamiltonian})
in the simplest case of single plasmonic mode and a set of identical
non-interacting two-level molecular emitters is

\begin{equation}
\mathrm{H}=\left[\begin{array}{cc}
E_{SPP} & V\\
V & E_{Exc}
\end{array}\right],\label{eq: Plexiton H matrix 2x2}
\end{equation}
which can be diagonalized analytically to yield the eigenvalues corresponding
to two new branches that arise due to the strong coupling between
SPP and exciton \cite{PhysRevLett.93.036404}: 

\begin{equation}
E_{\pm}=\frac{1}{2}\left(E_{Exc}+E_{SPP}\pm\sqrt{\left(E_{Exc}-E_{SPP}\right)^{2}+4V^{2}}\right).\label{eq: Plexiton branches}
\end{equation}

In the case of molecular emitters with doubly degenerate excited state,
such as F$_{n}$ZnPc in the minimal energy configuration, the Hamiltonian
matrix takes the form 

\begin{equation}
\mathrm{H}=\left[\begin{array}{ccc}
E_{SPP} & V_{01} & V_{02}\\
V_{01}^{*} & E_{Exc} & 0\\
V_{02}^{*} & 0 & E_{Exc}
\end{array}\right],\label{eq:Plexciton H matrix 3x3}
\end{equation}
where the coupling constants $V_{01}$ and $V_{02}$ account for different
spatial orientation of transition dipole moments of two degenerate
excitonic states $V_{0i}=g_{\boldsymbol{\mu}_{0i}}^{N}\left(\mathbf{k}\right)$,
$i$=1,2. The Hamiltonian matrix (\ref{eq:Plexciton H matrix 3x3})
can be analytically diagonalized as well. Two of it's eigenvalues
are similar to Eq. (\ref{eq: Plexiton branches})

\begin{equation}
E_{\pm}=\frac{1}{2}\left(E_{Exc}+E_{SPP}\pm\sqrt{\left(E_{Exc}-E_{SPP}\right)^{2}+4U^{2}}\right),\label{eq:plexciton branches 3x3}
\end{equation}
with coupling constant $U=\sqrt{V_{01}^{2}+V_{02}^{2}}$, and the
third one is $E_{0}=E_{Exc}$. 

\subsubsection{Ordered molecular aggregate}

In order to describe a coupled system consisting of a spatially ordered
molecular aggregate with surface plasmons we combine the formalism
described in Sec. A for the excitonic system with the quantum-mechanical
approach to describe plasmon-excitonic interaction \cite{GonzalesTudela2013,YuenZhou2018}
discussed above. The coupled Hamiltonian of a plasmon-excitonic system
(\ref{eq:Plexciton hamiltonian}) now reads

\begin{multline}
\mathrm{H}=\hbar\omega(\mathbf{k})a_{\mathbf{k}}^{\dagger}a_{\mathbf{k}}+\underset{{\scriptscriptstyle \alpha}}{\sum}\hbar\omega_{\alpha}D_{\alpha}^{\dagger}D_{\alpha}-\\
\underset{{\scriptscriptstyle I}}{\sum}\sqrt{\frac{\omega(\mathbf{k})}{2\epsilon_{0}AL(\mathbf{k})}}\left[a_{\mathbf{k}}e^{i\mathbf{k}\cdot\mathbf{R}_{I}-k_{z}z_{I}}\left(\mathbf{e}_{\mathbf{k}}+i\frac{\left|\mathbf{k}\right|}{k_{z}}\mathbf{e}_{z}\right)\cdot\hat{\boldsymbol{\mu}}_{I}+h.c.\right],\label{eq:Plexiton hamiltonian array}
\end{multline}
where index $\alpha$ runs over all excitonic states (\ref{eq:aggregate wavefunction})
of the molecular aggregate and $I$ over all individual molecular
emitters with dipole moments $\boldsymbol{\mu}_{I}$ located at points
$\left(\mathbf{R}_{I},z_{I}\right)$, $\mathbf{R}_{I}$ denoting an
in-plane and $z_{I}$ an out-of-plane position of an individual emitter.
It is essential to note, that the part of Hamiltonian describing the
interaction of SPP electric field with molecular emitters explicitly
includes the position of the emitter.

The matrix representation of the Hamiltonian (\ref{eq:Plexiton hamiltonian array})
has the general form

\begin{equation}
\mathrm{H}=\left[\begin{array}{cc}
E_{SPP} & \left[V\right]^{\dagger}\\
\left[V\right] & \left[E_{Exc}\right]
\end{array}\right],\label{eq:Plexciton H matrix array}
\end{equation}
where $\left[E_{Exc}\right]$ is a diagonal matrix with excitation
energies $\hbar\omega_{\alpha}$ standing on the diagonal, $\left[V\right]$
is a column containing the coupling between the particular excitonic
state and the SPP electric field:

\begin{equation}
V_{\alpha0}=\sqrt{\frac{\omega(\mathbf{k})}{2\epsilon_{0}AL(\mathbf{k})}}\left(\mathbf{e}_{\mathbf{k}}+i\frac{\left|\mathbf{k}\right|}{k_{z}}\mathbf{e}_{z}\right)\cdot\mathbf{M}_{\alpha0}\left(\mathbf{k}\right).\label{eq: Array plexciton coupling}
\end{equation}
Here, the collective excitonic transition dipole moments $\mathbf{M}_{\alpha0}\left(\mathbf{k}\right)$
are defined similarly to Eq. (\ref{eq:TDM}), but additionally include
the phase factor 

\begin{equation}
\mathbf{M}_{\alpha0}\left(\mathbf{k}\right)=\underset{{\scriptstyle {\scriptscriptstyle I}}}{\sum}\underset{{\scriptstyle i\ldots j\ldots z}}{\sum}\underset{{\scriptstyle j'}}{\sum}C_{i\dots j\dots z}^{*\alpha}C_{i\dots j'\dots z}^{0}e^{i\mathbf{k}\cdot\mathbf{R}_{I}-k_{z}z_{I}}\boldsymbol{\mu}_{jj'}^{I}.\label{eq: TDM array}
\end{equation}
Diagonalization of the matrix (\ref{eq:Plexciton H matrix array})
gives rise for multiple plexcitonic branches exhibiting several anti-crossings,
which can be related to different structural features of the molecular
aggregate.

\section{Computational details}

In order to obtain the excitation energies, transition dipole moments
and other quantities needed to perform the simulations according to
the method described above, the following calculations were carried
out in the frame of linear-response TDDFT. First, the nuclear geometries
of F$_{n}$ZnPc (n = 0, 4, 8, 16) were optimized using the long-range
dispersion-corrected functional $\omega$B97X-D \cite{Chai2008} and
6-311++G{*}{*} basis set for all atoms as implemented in Gaussian16
computational package \cite{g16}. For the optimized nuclear configurations,
the 2 lowest excited state energies and corresponding transition dipole
moments were calculated.

Dielectric function of a gold thin film needed to simulate the dispersion
of SPP was represented using the general spectral model with parameters
published in \cite{Ngoc2015}. Relative dielectric constant of self-assembled
monolayer was assumed to be 1.3.

\section{Results of simulations}

\subsection{Excited states in ordered F$_{n}$ZnPc aggregates}

\begin{figure}[H]
\includegraphics[width=0.9\textwidth]{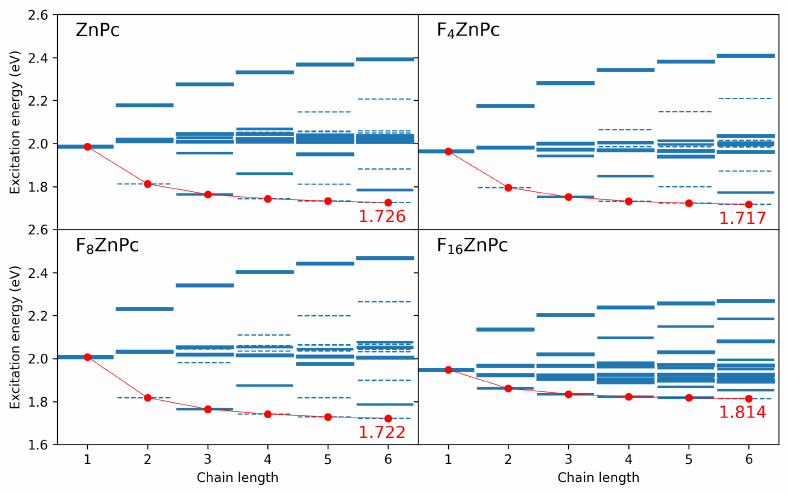}

\caption{Excited state energies of F$_{n}$ZnPc chains (Chain A) stacked in
the direction of the shortest ($\mathbf{a}$) unit cell vector (see
Fig. 1 d) in the main text). \label{fig:levels_X} }
\end{figure}

\begin{figure}[H]
\includegraphics[width=0.9\textwidth]{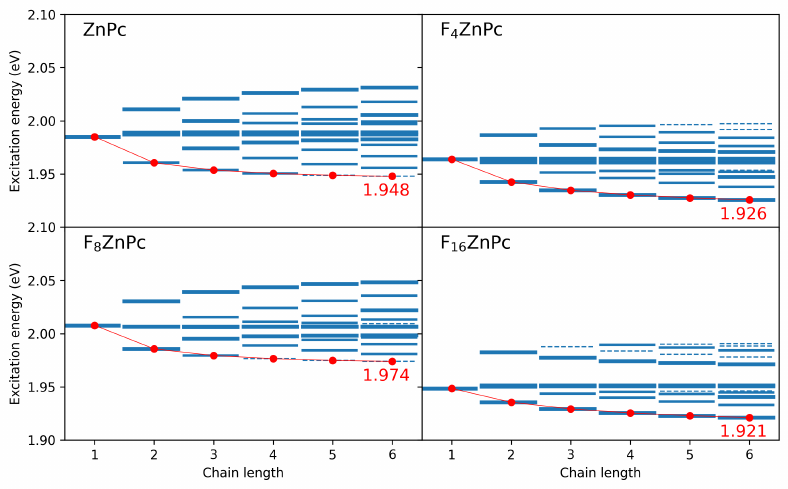}

\caption{Excited state energies of F$_{n}$ZnPc chains (Chain B) stacked in
the direction of the unit cell vector $\mathbf{b}$\label{fig:levels_Y}. }
\end{figure}

\begin{figure}[H]
\includegraphics[width=0.9\textwidth]{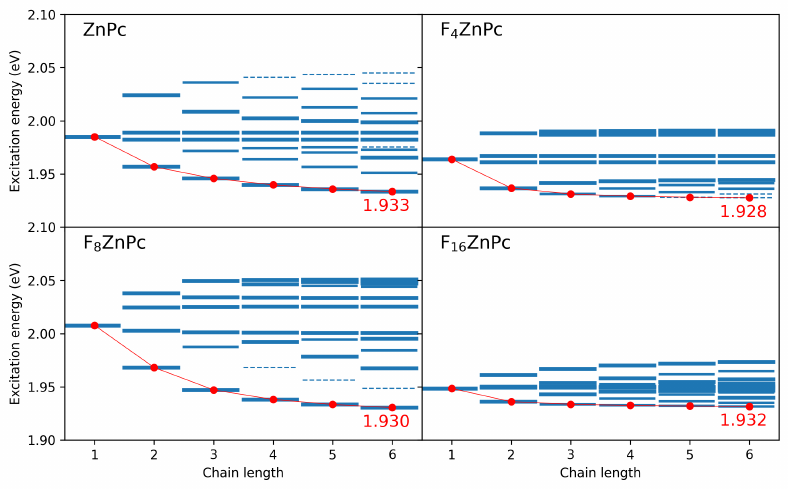}

\caption{Excited state energies of F$_{n}$ZnPc chains (Chain C) stacked in
the direction of the out-of-plane ($\mathbf{c}$) unit cell vector.
\label{fig:levels_Z} }
\end{figure}

In Figs. \ref{fig:levels_X} - \ref{fig:levels_Z} the excitation
energies of 1-dimensional F$_{n}$ZnPc chains built up in the directions
of $\mathbf{a}$ and $\mathbf{b}$ unit cell vectors (see Fig. 1d)
in the main text) assuming, that these vectors are parallel to the
metal-dielectric interface and the unit cell vector c, perpendicular to it. The energy of the lowest excited state
is in a very good agreement with the experimentally determined positions
of the anti-crossing 1 (Chain A, stacking in the direction of the
vector $\mathbf{a}$, Fig. \ref{fig:levels_X}) and the anti-crossing
2 (Chain B, stacking in the direction of the vector $\mathbf{b}$,
Fig. \ref{fig:levels_Y} and Chain C, stacking in the direction of
the vector $\mathbf{c}$, Fig. \ref{fig:levels_Z}). The excitation
energies were obtained via diagonalization of the Hamiltonian \ref{eq:total Hamiltonian}.
The unit cell parameters needed to construct the molecular chains
were taken from the crystal structure data of the F$_{n}$CuPc analogues
\cite{CuPc,F4F8CuPc,F16CuPc}.

\subsection{Anti-crossings in ordered F$_{n}$ZnPc aggregates}

In Table \ref{tab:en-coupl-chain} the energies and coupling strengths
calculated according to Eqs. \ref{eq:Plexciton H matrix array}-\ref{eq: TDM array}
at the lowest anti-crossings of the 1-dimensional chains A, B, and
C (chain length is 6 molecules) are presented.

\begin{table}[H]
\begin{tabular}{|c|c|c|c|c|c|c|}
\hline 
 & \multicolumn{2}{c|}{Chain A} & \multicolumn{2}{c|}{Chain B} & \multicolumn{2}{c|}{Chain C}\tabularnewline
\hline 
 & E (eV) & U (meV) & E (eV) & U (meV) & E (eV) & U (meV)\tabularnewline
\hline 
\hline 
ZnPc & 1.743 & 7.5 & 1.999 & 64.2 & 1.938 & 88.9\tabularnewline
\hline 
F$_{4}$ZnPc & 1.745 & 4.7 & 1.947 & 66.1 & 1.926 & 90.6\tabularnewline
\hline 
F$_{8}$ZnPc & 1.735 & 3.2 & 2.019 & 57.7 & 1.983 & 76.6\tabularnewline
\hline 
F$_{16}$ZnPc & 1.883 & 5.2 & 1.946 & 62.5 & 1.943 & 47.5\tabularnewline
\hline 
\end{tabular}

\caption{Energies and coupling strengths of anti-crossings in F$_{n}$ZnPc
chains A, B and C, constructed as described above. \label{tab:en-coupl-chain}}
\end{table}

\subsection{Anti-crossings in disordered F$_{n}$ZnPc layers}

In Table \ref{tab:en-coupl-monolayer} energies and coupling strengths
at anti-crossings in disordered mono- and double layers of F$_{n}$ZnPc,
calculated according to Eqs. (\ref{eq:Plexciton H matrix 3x3})-(\ref{eq:plexciton branches 3x3}).
In the layers I and II individual molecules are randomly oriented
within the whole solid angle 4$\pi$. In the layers III and IV molecules
are uniformly oriented with respect to the azimuthal angle (in-plane),
but the inclination angle from the metal-dielectric interface is uniformly
distributed in range {[}$\beta_{n}$-15$^{\mathrm{o}}$,$\beta_{n}$+15$^{\mathrm{o}}${]}.
The angles $\beta_{n}$ are assumed to increase with the degree of
fluorination and lie in the range of 60$^{\mathrm{o}}$-70$^{\mathrm{o}}$
as it is observed in crystalline structures F$_{n}$ZnPc and F$_{n}$CuPc
\cite{CuPc,F4F8CuPc,F16CuPc}. Particularly, in the current simulations
$\beta_{0}$=63$^{\mathrm{o}}$, $\beta_{4}$=65$^{\mathrm{o}}$,
$\beta_{8}$=68$^{\mathrm{o}}$, and $\beta_{16}$=70$^{\mathrm{o}}$.

\begin{table}[H]
\begin{tabular}{|c|c|c|c|c|c|c|c|c|}
\hline 
 & \multicolumn{2}{c|}{I} & \multicolumn{2}{c|}{II} & \multicolumn{2}{c|}{III} & \multicolumn{2}{c|}{IV}\tabularnewline
\hline 
 & E (eV) & U (meV) & E (eV) & U (meV) & E (eV) & U (meV) & E (eV) & U (meV)\tabularnewline
\hline 
\hline 
ZnPc & 1.979 & 41.35 & 1.973 & 58.06 & 1.976 & 51.04 & 1.967 & 71.46\tabularnewline
\hline 
F$_{4}$ZnPc & 1.958 & 41.50 & 1.952 & 58.22 & 1.954 & 54.98 & 1.943 & 76.88\tabularnewline
\hline 
F$_{8}$ZnPc & 2.002 & 41.06 & 1.996 & 57.74 & 1.995 & 61.94 & 1.982 & 86.42\tabularnewline
\hline 
F$_{16}$ZnPc & 1.942 & 41.38 & 1.937 & 58.08 & 1.932 & 69.98 & 1.915 & 97.30\tabularnewline
\hline 
\end{tabular}

\caption{Energies and coupling strengths of anti-crossings in disordered layers
of F$_{n}$ZnPc, I: uniform angular distribution, monolayer; II: uniform
angular distribution, double layer; III: uniform distribution in range
{[}$\beta_{n}$-15$^{\mathrm{o}}$,$\beta_{n}$+15$^{\mathrm{o}}${]},
monolayer; IV: uniform distribution in range {[}$\beta_{n}$-15$^{\mathrm{o}}$,$\beta_{n}$+15$^{\mathrm{o}}${]},
double layer. \label{tab:en-coupl-monolayer}}
\end{table}

\bibliography{SuppInfo}